\documentclass[a4paper,11pt]{article}
\pdfoutput=1

\usepackage{jheppub}

\usepackage{CJKutf8}
\usepackage{amsthm}
\usepackage{orcidlink}
\usepackage{lineno}

\hypersetup{hidelinks}
\numberwithin{equation}{section}

\theoremstyle{plain}
\newtheorem{theorem}{Theorem}[section]
\newtheorem{proposition}[theorem]{Proposition}
\newtheorem{lemma}[theorem]{Lemma}
\newtheorem{corollary}[theorem]{Corollary}

\theoremstyle{definition}

\theoremstyle{remark}

\title{A hyperbolicity obstruction near regular charged centers}

\author{Ye Zhou~(\begin{CJK*}{UTF8}{gbsn}周烨\end{CJK*})\,\orcidlink{0009-0004-7050-7736}}

\affiliation{Independent Researcher,\\
Kunshan, Jiangsu, China}
\emailAdd{ye.zhou.horizon@gmail.com}

\abstract{
A conserved charge creates a tension at a spherical center: the flux remains fixed while the area of the symmetry spheres shrinks. We ask whether curvature-dependent electromagnetic responses can reconcile bounded central curvature with hyperbolic propagation in four-dimensional Einstein gravity when both parity-even and parity-odd couplings are allowed. The background equations force a positive charge potential and its first curvature derivative to vanish at the central curvature endpoint. On branches with a positive electromagnetic kinetic coefficient and a positive effective gravitational coefficient, this endpoint flattening produces a physical extraordinary mode of the fully coupled equations whose squared angular characteristic speed is strictly below minus one at radii arbitrarily close to the center. Its characteristic cone has no hyperbolicity direction in common with the ordinary photon and graviton modes. The result requires no central curvature expansion, remains valid at points where the curvature Legendre Hessian vanishes, and extends to positive-definite mixing of finitely many Abelian fields. An exact analytic example with a nonlinear gravitational curvature sector admits a real, subluminal extraordinary photon cone near the center. In a fixed-action black-hole family, the obstruction extends outside the outer horizon over a near-extremal parameter interval.}
\arxivnumber{2611.24634}

\begin{document}
\maketitle
\flushbottom
\newcommand{\dd}{\mathrm{d}}
\newcommand{\cE}{\mathcal E}
\newcommand{\cC}{\mathcal C}

\section{Introduction}\label{Sec:Intro}
The area of a symmetry sphere shrinks to zero at a regular center, while a conserved electric displacement or magnetic flux remains fixed. With a constant Maxwell coupling, the resulting electromagnetic energy grows as $r^{-4}$. A charged center of bounded curvature therefore requires a markedly different response to its electromagnetic field. Nonlinear electrodynamics provides one approach, rooted in the constitutive theories of Born and Infeld and of Heisenberg and Euler~\cite{BornInfeld1934,HeisenbergEuler1936}. Regular black-hole geometries have also been explored since Bardeen's proposal, including the Dymnikova and Hayward models and charged constructions based on nonlinear sources~\cite{Bardeen1968,Dymnikova1992,Hayward2006,AyonBeatoGarcia1998,AyonBeatoGarcia2000,BronnikovComment2000,Bronnikov2001,FanWang2016,Ansoldi2008}; work on causal nonlinear electrodynamics has extended the discussion to charged AdS black holes, their thermodynamic phases and the matter dependence of their central singularity structure~\cite{BabaeiAghbolaghVelniHeIsapour2026}. These studies also highlight the distinction between bounded curvature, regular matter fields, geodesic completeness and a well-posed initial-value problem~\cite{Geroch1968,EllisSchmidt1977,Maeda2022}.

Another possibility is to let the electromagnetic response depend on spacetime curvature. Such interactions appear in early studies of nonminimal electrodynamics, in quantum-electrodynamic effective actions and dimensional reduction, and in cosmological applications of nonminimal Maxwell--$F(R)$ couplings, including inflation, late-time acceleration and large-scale magnetic-field generation~\cite{Prasanna1971,Horndeski1976,Buchdahl1979,DrummondHathrell1980,MullerHoissen1988,MullerHoissenSippel1988,Balakin2005,BambaOdintsov2008}. The scalar-curvature term $h(R)F_{ab}F^{ab}$ is particularly useful because its coupling can be related directly to the geometry and the conserved charge. Exact solutions, screening mechanisms and compact-star configurations have been investigated~\cite{DereliSert2011Log,DereliSert2011Power,Sert2012,Sert2013,Sert2016,Sert2018Stars,Sert2020,Dereli2020}. More general actions incorporate curvature-tensor couplings, electromagnetic duality and higher-derivative interactions~\cite{Balakin2008,CanoMurcia2020,Cano2021,CanoMurciaDuality2021,CanoMurciaExact2021,Colleaux2024}. Recent work by Bokuli\'c and collaborators derives trace-based restrictions on regular charged backgrounds in several effective-field-theory-motivated curvature-coupling sectors, including interactions involving the pseudoscalar electromagnetic invariant. Their analysis also identifies simultaneous curvature couplings to both electromagnetic invariants as a natural but nontrivial extension~\cite{Bokulic2026}.

Even when a regular charged background exists, a different question remains: can its perturbations be evolved from suitable initial data? The connection between electromagnetic constitutive laws and characteristic cones is well understood~\cite{Boillat1970,Novello2000,GibbonsHerdeiro2001,Abalos2015}. De Felice and Tsujikawa, in particular, found an angular Laplacian instability in regular black holes supported by nonlinear electrodynamics and studied the effect of additional scalar couplings~\cite{DeFeliceTsujikawa2025,DeFeliceTsujikawaScalar2025}. Curvature-dependent Maxwell responses introduce a different mechanism. Whereas nonlinear electrodynamics modifies the constitutive dependence on electromagnetic invariants, a curvature-dependent coefficient couples the electromagnetic perturbation to the scalar curvature and hence to the gravitational constraints. Its physical characteristic cone must therefore be obtained from the coupled system. The question addressed here is complementary to the background-existence restrictions just mentioned: once a charged background satisfies the field equations, do all of its physical modes admit a common Cauchy direction? Related questions of symmetric hyperbolicity and gravitational principal symbols have been studied in Refs.~\cite{Goulart2022,ChenStability2024,PapalloReall2017,KovacsReall2020}.

We consider four-dimensional Einstein gravity with a parity-even response $h(R)$ and a parity-odd response $\theta(R)$. For positive electromagnetic response, the charges combine into a positive curvature-dependent potential $\mathcal C(R;P,Q)$. We first show that the background field equations force both $\mathcal C$ and $\mathcal C_R$ to vanish at the central curvature endpoint. This condition does more than control the background stress: it makes the extraordinary angular characteristic of the coupled perturbations negative, with $c_\perp^2<-1$ at radii arbitrarily close to the center whenever the electromagnetic response and effective gravitational coefficient are positive. The extraordinary and ordinary characteristic quadratics then have disjoint hyperbolicity cones. The proof uses the bounded-curvature equations directly and includes points where the curvature Legendre Hessian vanishes.

The obstruction extends to positive-definite mixing among finitely many Abelian fields. An analytic $\Phi(R)$ example shows that adding a gravitational curvature Hessian can instead restore a real, subluminal extraordinary photon cone near the center, and an exact family with a fixed electromagnetic action places the obstruction outside the outer horizon over a near-extremal interval. We develop the central endpoint identities first, derive the physical characteristics and their hyperbolicity directions, and then examine these extensions and black-hole geometries.

\section{Charged centers}\label{Sec:Centers}

Use signature $(-+++)$ and the curvature convention in which de Sitter space has positive scalar curvature. Let
\begin{equation}
\begin{aligned}
 S&=\frac1{16\pi}\int\sqrt{-g}\,[R-2\Lambda-h(R)U-\theta(R)G]\,\dd^4x,\\
 U&=F_{ab}F^{ab},\qquad G=F_{ab}{}^\star F^{ab}.
\end{aligned}
\label{eq:action}
\end{equation}
Our orientation is fixed by $U=2(b^2-e^2)$ and $G=-4eb$ for $F=-e\,\omega^0\wedge\omega^1+b\,\omega^2\wedge\omega^3$. In this convention the closed displacement form is $\mathcal D=-h\,{}^\star F+\theta F$. Define
\begin{equation}
 V=-h_RU-\theta_RG,\qquad W=1+V.\label{eq:VW}
\end{equation}
At fixed covariant $F_{ab}$, $\delta U=2F_{ac}F_b{}^c\delta g^{ab}$ and $\delta G=Gg_{ab}\delta g^{ab}/2$. The latter identity follows from metric independence of $\sqrt{-g}G$. It cancels the algebraic odd-response stress. Integrating the $W\delta R$ term by parts gives
\begin{align}
 G_{ab}+\Lambda g_{ab}
 &=2h\left(F_{ac}F_b{}^c-\tfrac14g_{ab}U\right)
   +\nabla_a\nabla_bV-g_{ab}\Box V-R_{ab}V,
 \label{eq:einstein}\\
 \dd F&=0,\qquad \dd\mathcal D=0.\label{eq:maxwell}
\end{align}
Taking compactly supported variations removes the boundary divergences from the $W\delta R$ term and the electromagnetic potential. The trace of the metric equation is
\begin{equation}
 R-4\Lambda=RV+3\Box V.\label{eq:trace}
\end{equation}

In a punctured spherical neighborhood write
\begin{align}
 \dd s^2&=-A(r)\dd t^2+\frac{\dd r^2}{B(r)}+r^2\dd\Omega_2^2,\label{eq:metric}\\
 F&=-\sqrt{A/B}\,e(r)\,\dd t\wedge\dd r
       +P\sin\vartheta\,\dd\vartheta\wedge\dd\varphi,\nonumber\\
 b&=P/r^2,\qquad Q=r^2h e+\theta P.\label{eq:charges}
\end{align}
Initially $A$ and $B$ are nonzero and have the same sign. We assume $A,B\in C^2$, $h,\theta\in C^1$ on an open interval containing the attained curvatures, $V\in C^2$, and a locally continuous real electromagnetic field with the derivatives required by Maxwell's equations. The classical equations hold at every positive radius. For the perturbation analysis the background fields and response laws are smooth at such points, where the principal Hessians involve two curvature derivatives. The limiting curvature may lie at the boundary of the response domain. In particular, bounded central curvature is assumed without a Taylor expansion or a smooth extension across the filled center.

The exact identity
\begin{align}
 K={}&\frac{4(1-B)^2}{r^4}+\frac{2B'^2}{r^2}
 +\frac{2B^2A'^2}{r^2A^2}
 +\left[\frac{AA'B'-B(A'^2-2AA'')}{2A^2}\right]^2
 \label{eq:Kexact}
\end{align}
is a nonnegative sum of squares in this symmetry class. Therefore $K=O(1)$ implies
\begin{equation}
\begin{gathered}
 B=1+O(r^2),\qquad B'=O(r),\qquad A'/A=O(r),\\
 A''/A=O(1),\qquad A\longrightarrow A_0>0.
\end{gathered}
\label{eq:metricbounds}
\end{equation}
The center therefore lies in a static region, and its Ricci components are bounded. The sum-of-squares identity gives these estimates directly, including in situations where comparisons among scalar curvature invariants require care~\cite{Dragasevic2025,Szybka2026}. The independent Einstein components are
\begin{align}
 G^t{}_t&=(B-1)/r^2+B'/r,\nonumber\\
 G^r{}_r&=(B-1)/r^2+BA'/(rA),\nonumber\\
 G^\vartheta{}_\vartheta&=B\left[\frac{A''}{2A}-\frac{A'^2}{4A^2}
 +\frac{A'B'}{4AB}+\frac{A'}{2Ar}+\frac{B'}{2Br}\right],
 \qquad R=-G^t{}_t-G^r{}_r-2G^\vartheta{}_\vartheta.
 \label{eq:components}
\end{align}

\begin{theorem}[Bounded composite response and source]\label{thm:load}
Under these background assumptions and $K=O(1)$,
\begin{align}
 V&=V_0+O(r^2),\quad V'=O(r),\quad V''=O(1),\quad\Box V=O(1),\label{eq:Vbounds}\\
 \cE&:=|h|(e^2+b^2)=O(1).\label{eq:load}
\end{align}
Neither positivity nor division by $h$ is required.
\end{theorem}
\begin{proof}
Set $p=(A'/A+B'/B)/2$ and $q=(A'/A-B'/B)/2$, both $O(r)$. The radial minus temporal Ricci and Hessian differences are respectively $2Bq/r$ and $B(V''-qV')$. The electromagnetic terms cancel exactly, leaving
\begin{equation}
 W''-qW'-\frac{2q}{r}W=0.\label{eq:Wode}
\end{equation}
The first-order system for $(W,W')$ has bounded coefficients. Backwards Gronwall control from any fixed positive radius bounds $W,W'$ and then $W''$. Equation \eqref{eq:trace} bounds $\Box V$, whereas $\Box V=B[V''+(2/r+p)V']$. Consequently $V'=O(r)$ and integration gives the asserted endpoint value. Finally,
\begin{equation}
 G^t{}_t+\Lambda=-h(e^2+b^2)+\frac{BA'}{2A}V'-\Box V-R^t{}_tV
 \label{eq:timeeq}
\end{equation}
bounds the absolute value of the source. This argument also holds at response zeros, where the algebraic source vanishes.
\end{proof}

For constant $\theta$, the charge algebra reduces to the parity-even theory after the shift $Q\mapsto Q-\theta P$. The bound $P^2|h|+(Q-\theta P)^2/|h|\le Cr^4$ on $h\ne0$ then excludes simultaneous nonzero $P$ and $Q-\theta P$. On a screened region, the undivided displacement law remains the appropriate description. Variable $\theta(R)$, by contrast, permits both charges to appear in the same solution.

The geometric regularity assumed here concerns the punctured neighborhood. Smooth extension of $F$ and of the displacement form, geodesic completeness, and smoothness of the metric in Cartesian coordinates are stronger requirements~\cite{EllisSchmidt1977,Olmo2016,BalartVagenas2014,Maeda2022,LimaJunior2025,Antonelli2026}. For example, Stokes' theorem forces $P=0$ if $F$ extends as a smooth closed two-form over the filled ball, and likewise forces $Q=0$ for a smooth displacement form. Screened and stealth sectors further illustrate the distinction between a field and its conserved flux~\cite{Smolic2018}.

We now combine the charges into a single curvature response. Theorem~\ref{thm:load} supplies the necessary source bounds even at response zeros; the electric field remains locally continuous at positive radius.

For fixed charges define, wherever $h\ne0$, the signed charge potential
\begin{equation}
 \mathcal C(R;Q,P)
 =P^2h(R)+\frac{[Q-P\theta(R)]^2}{h(R)}.
 \label{eq:chargepotential}
\end{equation}
The two terms have the sign of $h$. Direct differentiation, using the undivided flux law, gives
\begin{equation}
 \mathcal E=\frac{|\mathcal C|}{r^4},
 \qquad
 \mathcal C_R(R(r))=-\frac{r^4}{2}V(r).
 \label{eq:chargepotentialidentity}
\end{equation}
Both curvature derivatives in \eqref{eq:VW} are necessary for the second identity.

\begin{theorem}[Joint endpoint bound]
\label{thm:jointendpoint}
Let a static, spherically symmetric solution of \eqref{eq:action} have bounded Kretschmann curvature and $(P,Q)\ne(0,0)$. Assume $h,\theta\in C^1$ on the attained curvature interval and the regularity needed for \eqref{eq:Vbounds}. Then $R(r)\to R_0$ for some finite $R_0$. If $M(r)=\sup_{0<s\le r}|V(s)|$, then
\begin{equation}
 \mathcal E(r)\le\frac{M(r)}2|R(r)-R_0|.
 \label{eq:jointmodulus}
\end{equation}
In particular, $\mathcal E=o(1)$. For $P\ne0$,
\begin{equation}
 h(R(r))=o(r^4),\qquad
 Q-P\theta(R(r))=o(r^4).
 \label{eq:jointstrict}
\end{equation}
For $P=0$, $Q\ne0$, one has $r^4|h|\to\infty$.
\end{theorem}

\begin{proof}
At a nonzero-response point the bounded source gives
\begin{equation}
 P^2|h|+\frac{(Q-P\theta)^2}{|h|}=r^4\mathcal E.
 \label{eq:jointload}
\end{equation}
At a response zero, \eqref{eq:charges} instead gives $Q=P\theta$, and $\mathcal E=0$.

When $P=0$, $Q\ne0$, the response never vanishes and $|h|\to\infty$. If $R$ had distinct accumulation values, every intermediate curvature would be attained along a sequence approaching the center. The finite value of $h$ at that curvature would then have to diverge. Hence $R$ has a limit.

When $P\ne0$, \eqref{eq:jointload} gives $h\to0$ and, by its arithmetic--geometric-mean bound, $\theta\to Q/P$. A nontrivial interval of curvature accumulation values would therefore satisfy $h=0$ and $\theta=Q/P$ identically. On its radial preimage both curvature derivatives vanish, so $V=0$ and the trace forces $R=4\Lambda$. This contradicts an interval of attained curvature values. Thus $R\to R_0$ in this case also.

For the magnetic case, extend $\mathcal C$ to every attained zero of $h$ by $\mathcal C=0$. This extension is continuous, including at nonisolated zeros. To see this, let $c_n\to c$ with $h(c)=0$, and choose $s_n\in(0,r]$ such that $R(s_n)=c_n$. After passing to a subsequence, either $s_n$ approaches a positive radius or it approaches zero. In the first case local boundedness of $e$ and
\begin{equation}
 \mathcal C(R(s_n))
 =s_n^4h(R(s_n))\left[e(s_n)^2+\frac{P^2}{s_n^4}\right]
 \label{eq:Ccontinuity}
\end{equation}
give convergence to zero. In the second case the bounded-source estimate gives $|\mathcal C(R(s_n))|=O(s_n^4)\to0$.

Every curvature strictly between $R_0$ and $R(r)$ occurs at some $s\le r$. Equation~\eqref{eq:chargepotentialidentity} therefore bounds $|\mathcal C_R|$ by $r^4M(r)/2$ off the response zero set. A continuous function vanishing on a closed set and with derivative bounded by $L$ on its complement is $L$-Lipschitz on the full interval. Indeed, use the mean-value theorem within a complementary component; for points separated by zeros, insert the nearest intervening zeros and add the two distance bounds. This proves the required estimate across arbitrary response zero sets. The endpoint value is zero because $\mathcal C(R(s))=O(s^4)$ as $s\to0$. Integration gives
\begin{equation}
 |\mathcal C(R(r))|
 \le\frac{r^4M(r)}2|R(r)-R_0|,
\end{equation}
which proves \eqref{eq:jointmodulus}. The electric case has no response zeros and the same integration applies directly.

Finally, \eqref{eq:jointload} implies $P^2|h|\le r^4\mathcal E$ and $2|P||Q-P\theta|\le r^4\mathcal E$. These yield \eqref{eq:jointstrict}. For the purely electric charged sector, $\mathcal E=Q^2/(r^4|h|)\to0$.
\end{proof}

If $|R-R_0|=O(r^\alpha)$, the two quantities in \eqref{eq:jointstrict} are $O(r^{4+\alpha})$. For positive response, the joint bound also has the geometric form
\begin{equation}
 \frac{(\theta-Q/P)^2+h^2}{h}
 \le\frac{r^4M(r)}{2P^2}|R-R_0|.
 \label{eq:responsecircle}
\end{equation}
It confines the complex response $\theta+ih$ to disks tangent to the real boundary point $Q/P$. This geometric form is familiar from electric--magnetic response matrices~\cite{GaillardZumino1981,GibbonsRasheed1995}; it does not imply a duality symmetry of a fixed action.

\begin{corollary}[Coupled orders of vanishing]
\label{cor:jointorders}
Let $P\ne0$, choose an attained one-sided coordinate $x=\pm(R-R_0)>0$, and suppose
\begin{equation}
 h=a x^k(1+o(1)),\qquad
 \theta-Q/P=b x^\ell(1+o(1)),\qquad ab\ne0,
 \label{eq:jointorders}
\end{equation}
with the corresponding first differentiated asymptotics. A bounded-curvature center requires
\begin{equation}
 \nu:=\min\{k,2\ell-k\}>1.
 \label{eq:nucondition}
\end{equation}
Moreover,
\begin{equation}
 x=O(r^{4/(\nu-1)}),\quad
 h=O(r^{4k/(\nu-1)}),\quad
 \theta-Q/P=O(r^{4\ell/(\nu-1)}).
 \label{eq:jointrates}
\end{equation}
\end{corollary}

\begin{proof}
The terms in $\mathcal C/P^2=h+(\theta-Q/P)^2/h$ have the same sign, so their leading coefficients cannot cancel. Its effective order is $\nu$. Its value tends to zero and its derivative tends to zero by \eqref{eq:chargepotentialidentity}; hence $\nu>1$. The derivative lower bound $|\mathcal C_R|\asymp x^{\nu-1}=O(r^4)$ proves the rates.
\end{proof}

For example, a cubic zero of $h$ with only quadratic matching of $\theta$ has $\nu=1$ and is excluded. Controlling the even response alone is insufficient.

The endpoint bounds are attained by exact solutions, including configurations with both electric and magnetic fields.

\begin{proposition}[An arbitrary-function lift]
\label{prop:jointlift}
Let a pure magnetic seed solution have charge $P\ne0$ and response $H(R)$. For any sufficiently differentiable real function $\kappa(R)$ and constant $\theta_0$, define
\begin{equation}
 h=\frac{H}{1+\kappa^2},\qquad
 \theta=\theta_0-\frac{\kappa H}{1+\kappa^2}.
 \label{eq:jointlift}
\end{equation}
The same geometry is an exact solution of the resulting action with
\begin{equation}
 b=P/r^2,\qquad e=\kappa(R(r))P/r^2,\qquad Q=\theta_0P.
 \label{eq:liftfields}
\end{equation}
\end{proposition}

\begin{proof}
Substitution gives
\begin{align}
 r^2he+\theta P&=\theta_0P,\qquad
 h(e^2+b^2)=HP^2/r^4,\notag\\
 -h_RU-\theta_RG&=-2P^2H_R/r^4.
 \label{eq:liftsources}
\end{align}
The Maxwell flux, signed Maxwell stress and complete composite curvature response are exactly those of the seed. Regular magnetic configurations with nonminimal curvature couplings have earlier precedents~\cite{BalakinZayats2015,Balakin2009}. Thus every gravitational component, including the angular equation, is preserved. The magnetic flux is constant and a static radial electric profile is closed, so both Maxwell equations hold.
\end{proof}

The lift produces a new response law with the same geometry and source terms as the magnetic seed. The resulting solutions test the joint endpoint bounds in genuinely electric--magnetic configurations, whose geometric regularity can coexist with a singular field strength at the filled center.

For the following polynomial lifts we set $\Lambda=0$. A concrete smooth geometric center is obtained from fixed parameters $R_0$, $\beta>0$, $\eta\ne0$ and $\theta_0$:
\begin{equation}
 x=R_0-R,\qquad
 h=\frac{\beta x^3}{1+\eta^2x^2},\qquad
 \theta=\theta_0-\frac{\beta\eta x^4}{1+\eta^2x^2}.
 \label{eq:cubicliftlaw}
\end{equation}
For a nonzero magnetic charge choose $c>0$ subject to
\begin{equation}
 5400\beta c^2P^2=1.
 \label{eq:cubicliftfixed}
\end{equation}
Then
\begin{align}
 A=B=N&=1-\frac{R_0}{12}r^2+cr^4,
 &R&=R_0-30cr^2,\notag\\
 b&=P/r^2,
 &e&=30\eta cP,
 &Q&=\theta_0P.
 \label{eq:cubicliftbackground}
\end{align}
Here $V=1$, $W=2$, and $h(e^2+b^2)=5cr^2$. For a one-function metric,
\begin{align}
 R^t{}_t=R^r{}_r&=-N''/2-N'/r,
 &R^\vartheta{}_\vartheta&=(1-N-rN')/r^2,\notag\\
 G^t{}_t=G^r{}_r&=(rN'+N-1)/r^2,
 &G^\vartheta{}_\vartheta&=N''/2+N'/r.
 \label{eq:liftRicci}
\end{align}
Substitution gives $G^t{}_t+R^t{}_tV+5cr^2=0$ and $G^\vartheta{}_\vartheta+R^\vartheta{}_\vartheta V-5cr^2=0$; the radial, trace and Maxwell equations vanish as well. These identities check the complete set of background equations. Relation~\eqref{eq:cubicliftfixed} fixes the allowed solution parameters for the response law~\eqref{eq:cubicliftlaw}. The geometry is locally smooth; its nonzero magnetic flux precludes a smooth source-free field extension over the filled ball.

The example is not a constant-$\theta$ relabeling. To determine whether a constant electromagnetic duality transformation of the closed doublet can make it purely magnetic, consider a nonzero constant linear combination $F'=uF+v\mathcal D$. Its electric part vanishes only if
\begin{equation}
 vh=(u+v\theta)\kappa.
 \label{eq:constantdualitycondition}
\end{equation}
On a branch with $\kappa\ne0$, this requires $v\ne0$ and, using \eqref{eq:jointlift},
\begin{equation}
 H/\kappa=\theta_0+u/v=\text{constant}.
 \label{eq:constantdualityobstruction}
\end{equation}
Thus a constant transformation can remove the electric field only when $H$ and $\kappa$ are proportional with a constant proportionality factor. The separate branch $\kappa\equiv0$ is already purely magnetic, since $e=\kappa b$ vanishes identically, and requires no duality transformation. But $H/\kappa=(\beta/\eta)x^2$ varies with $x$ for the cubic lift in \eqref{eq:cubicliftlaw}. Thus no constant invertible transformation of the closed doublet makes this genuinely electric--magnetic configuration purely magnetic. By contrast, constant $\theta$ merely shifts $Q$ by $\theta P$ and adds no such example.

\section{The hyperbolicity obstruction}\label{Sec:Hyperbolicity}

We now turn from the background response to the physical characteristics of the coupled gravitational and electromagnetic equations. Throughout this section the coefficients are evaluated at a smooth positive-radius point.

\subsection{The coupled characteristic polynomial}

Write
\begin{equation}
 f(R,U,G)=R-2\Lambda-h(R)U-\theta(R)G,
 \qquad T=f_{RR}=-h_{RR}U-\theta_{RR}G.
 \label{eq:curvatureHessian}
\end{equation}
When $T\ne0$, the Legendre variable $W=f_R$ determines $R=\chi(W,U,G)$ locally. The equivalent action is $WR-\mathcal P(W,U,G)$ with $\mathcal P=W\chi-f(\chi,U,G)$. At fixed $W$,
\begin{equation}
 \mathcal P_U=h,\quad\mathcal P_G=\theta,\qquad
 \begin{pmatrix}\mathcal P_{UU}&\mathcal P_{UG}\\
 \mathcal P_{GU}&\mathcal P_{GG}\end{pmatrix}
 =\frac1T
 \begin{pmatrix}h_R^2&h_R\theta_R\\h_R\theta_R&\theta_R^2\end{pmatrix}.
 \label{eq:LegendreHessian}
\end{equation}
For $W>0$, the conformal metric $\widetilde g_{ab}=Wg_{ab}$ puts gravity in Einstein form and produces the scalar kinetic term $-3(\widetilde\nabla\log W)^2/2$. Scalar--photon mixing from the potential has lower differential order than the kinetic blocks. The same is true of metric--photon mixing after Einstein gauge fixing. The photon characteristic block is consequently part of the full coupled principal system. This distinction is also central to gravitational well-posedness analyses~\cite{PapalloReall2017,KovacsReall2020}.

Take the background fields along the local $z$ axis and define
\begin{equation}
 \alpha=h_Re+\theta_Rb,\qquad
 \beta_1=h_Rb-\theta_Re.
 \label{eq:principalalpha}
\end{equation}
The photon quadratic Lagrangian, up to an overall positive normalization, is
\begin{equation}
 \mathcal L^{(2)}_{\rm ph}
 =2h(\delta\boldsymbol E^2-\delta\boldsymbol B^2)
 +4\theta\delta\boldsymbol E\mathbin{\cdot}\delta\boldsymbol B
 -\frac8T(\beta_1\delta B_z-\alpha\delta E_z)^2.
 \label{eq:photonquadratic}
\end{equation}
The frozen $\theta$ term is a total derivative. For a wave along $x$, the Gauss constraint sets the longitudinal electric perturbation to zero; the two transverse potentials are $A_y,A_z$. Their determinant is proportional to
\begin{equation}
 h(\omega^2-k^2)
 \left[\left(h-\frac{4\alpha^2}{T}\right)\omega^2
 -\left(h+\frac{4\beta_1^2}{T}\right)k^2\right].
 \label{eq:photoncharacteristic}
\end{equation}
Thus one polarization is luminal. Necessary signs on this nondegenerate Legendre branch are
\begin{equation}
 h>0,\qquad
 \mathcal K=h-4\alpha^2/T>0,\qquad
 \mathcal G_\perp=h+4\beta_1^2/T>0.
 \label{eq:photonsigns}
\end{equation}
These are local necessary sign conditions on the nondegenerate Legendre branch.

To include $T=0$, retain both auxiliary fields instead of eliminating curvature:
\begin{equation}
 \mathcal L_{\rm aux}=W(R-\chi)+f(\chi,U,G).
 \label{eq:uneliminatedaction}
\end{equation}
Variation of $W$ gives $\chi=R$, so this action is exactly equivalent even at $T=0$. Its Einstein-frame potential contains
\begin{equation}
 \frac{(1-W)\chi-2\Lambda}{W^2}
 -h(\chi)\widetilde U-\theta(\chi)\widetilde G.
 \label{eq:auxEFpotential}
\end{equation}
Let $\varphi=\delta W$, $z=\delta\chi$, $\gamma=3/W^2$, and $m=(W-2)/W^3$. Denote Einstein-frame electromagnetic coefficients by $\widehat\alpha=\alpha/W$, $\widehat\beta_1=\beta_1/W$, $\widehat T=T/W^2$. The relevant local quadratic terms are
\begin{align}
 \mathcal L^{(2)}_{
 \rm aux}={}&2h(\dot A_y^2+\dot A_z^2-A_{y,x}^2-A_{z,x}^2)
 -4z(\widehat\beta_1 A_{y,x}+\widehat\alpha\dot A_z)\notag\\
 &+\frac{\widehat T}{2}z^2+mz\varphi
 +\frac\gamma2(\dot\varphi^2-\varphi_{,x}^2),
 \label{eq:auxquadratic}
\end{align}
apart from scalar potential terms and background gradients that do not affect the non-null high-frequency root below. For $\Delta=\omega^2-k^2$, the Fourier matrix is
\begin{equation}
 \begin{pmatrix}
 h\Delta&0&0&i\widehat\beta_1k\\
 0&h\Delta&0&-i\widehat\alpha\omega\\
 0&0&\gamma\Delta&m\\
 -4i\widehat\beta_1k&4i\widehat\alpha\omega&m&\widehat T
 \end{pmatrix}.
 \label{eq:auxmatrix}
\end{equation}
Its determinant is
\begin{equation}
 h\Delta^2
 \left\{\gamma[(h\widehat T-4\widehat\alpha^2)\omega^2
 -(h\widehat T+4\widehat\beta_1^2)k^2]-hm^2\right\}.
 \label{eq:auxdet}
\end{equation}
The $m^2$ term is lower order in frequency. On a non-null root $\Delta$ is of order $k^2$, and the scalar equation gives $\varphi=-mz/(\gamma\Delta)$; scalar feedback into the auxiliary constraint is therefore suppressed by $k^{-2}$. This also explains why the constraint must be kept when $\widehat T=0$. Removing the common positive conformal factors, the extraordinary high-frequency polynomial is
\begin{equation}
 L\omega^2-Mk^2,
 \qquad L=hT-4\alpha^2,\quad M=hT+4\beta_1^2.
 \label{eq:universalpolynomial}
\end{equation}
At $T=0$ with $\alpha\beta_1\ne0$, one has $L=-4\alpha^2$ and $M=4\beta_1^2$, so the extraordinary squared speed is $-\beta_1^2/\alpha^2$. This root follows directly from the uneliminated auxiliary constraint and does not require division by $T$. The rank changes when $\alpha=0$; the endpoint argument below selects witnesses on the nondegenerate extraordinary branch.

The physical polarization can be exhibited on the non-null branch. For $q=\lambda\widehat q$ with $\widehat q^2\ne0$, take $z=O(1)$. The transverse Maxwell amplitudes are $O(\lambda^{-1})$, while the harmonic-gauge metric and propagating scalar amplitudes are $O(\lambda^{-2})$, owing to their invertible non-null wave blocks. Choosing $z\ne0$, the leading transverse amplitudes are
\begin{equation}
 A_y=-\frac{i\widehat\beta_1k}{h\Delta}z,\qquad
 A_z=\frac{i\widehat\alpha\omega}{h\Delta}z,\qquad
 \varphi=-\frac{m}{\gamma\Delta}z.
 \label{eq:physicalnonnullvector}
\end{equation}
The last row of \eqref{eq:auxmatrix} yields the extraordinary dispersion relation. Appendix~\ref{App:PrincipalReduction} carries out the Schur reduction in the full harmonic-gauge Einstein system, identifies the one-dimensional physical kernel, and verifies the gravitational and electromagnetic subsidiary constraints. It also establishes persistence of a simple non-null physical root after restoration of the frozen lower-order terms.

For $P\ne0$ and $h>0$, introduce
\begin{equation}
 \kappa=\frac{Q/P-\theta}{h}=e/b,
 \qquad H=h(1+\kappa^2)=\mathcal C/P^2>0.
 \label{eq:effectiveH}
\end{equation}
Primes in the remainder of this section denote curvature derivatives. Direct differentiation gives
\begin{align}
 \alpha&=-bh\kappa',&
 \beta_1&=b(H'-h\kappa\kappa'),\notag\\
 T&=-2b^2(H''-2h\kappa'^2),&
 L&=-2b^2hH''.
 \label{eq:Heffectiveidentities}
\end{align}
Whenever $H''\ne0$, the squared speed is
\begin{equation}
 c_\perp^2=\frac ML
 =1-\frac{2H'^2}{HH''}
 -\frac{2H}{H''}
 \left(\kappa'-\kappa\frac{H'}H\right)^2.
 \label{eq:Hspeed}
\end{equation}
This formula follows from the uneliminated polynomial and remains valid at $T=0$. On $T\ne0$, one also has
\begin{equation}
 \mathcal K=\frac{hH''}{H''-2h\kappa'^2}.
 \label{eq:Hkinetic}
\end{equation}

The angular calculation has a covariant form that also identifies the physical polarization. At a smooth positive-radius point define
\begin{equation}
 J^{ab}=h_R F^{ab}+\theta_R{}^\star F^{ab},\qquad
 j^b=q_aJ^{ab},\qquad q^2=g^{ab}q_aq_b.
 \label{eq:covariant-J}
\end{equation}
For a non-null Fourier covector, choose Maxwell Lorenz gauge and write $z=\delta\chi$. The leading Maxwell and auxiliary equations are
\begin{equation}
 hq^2a^b-ij^bz=0,\qquad Tz-4ij^ba_b=0.
 \label{eq:covariant-system}
\end{equation}
The first equation retains all four components, including Gauss's law. Antisymmetry gives $q_bj^b=0$. Thus
\begin{equation}
 a^b=\frac{ij^b}{hq^2}z,\qquad
 \mathcal P(q):=hTq^2+4j^2=0.
 \label{eq:covariant-root}
\end{equation}
The same non-null polynomial applies at $T=0$. At the strictly negative roots established below, the field-strength amplitude is nonzero. The ordinary branch $q^2=0$ carries a photon polarization with $z=0$ and $j\cdot a=0$, after the Maxwell gauge quotient.

In an orthonormal frame adapted to the spherical orbits, $J^{03}=\alpha$ and $J^{12}=\beta_1$. If $k_s$ denotes the spatial component in the Lorentzian orbit plane and $k_\perp$ the angular magnitude,
\begin{align}
 j^2&=\alpha^2(\omega^2-k_s^2)+\beta_1^2k_\perp^2,\notag\\
 \mathcal P(q)&=-L(\omega^2-k_s^2)+Mk_\perp^2.
 \label{eq:full-covector}
\end{align}
The direction $s$ is radial in a static region. The orbit-plane formulation remains meaningful in a trapped region without treating a spacelike static-coordinate vector as time.

Antisymmetry gives $q\cdot j=0$, so the non-null Maxwell solution satisfies the longitudinal Maxwell equation. At an extraordinary root, $j\ne0$ and $a^b$ is proportional to $j^b$; because $q^2\ne0$ and $q\cdot j=0$, this amplitude cannot be proportional to $q^b$. Hence $q_{[a}a_{b]}\ne0$: the mode carries a nonzero principal field strength and is not a pure Maxwell-gauge perturbation. This conclusion also holds at $T=0$, where the retained auxiliary constraint imposes $j^2=0$ rather than removing the mode. The auxiliary formulation remains equivalent to the original field equations because the $W$ equation enforces $\chi=R$. Appendix~\ref{App:PrincipalReduction} verifies the gravitational and electromagnetic gauge constraints for this polarization.

\subsection{Flat endpoints and angular propagation}
The endpoint estimate controls not only the response but also its curvature derivative. The following elementary lemma turns that information into a characteristic obstruction without a Taylor expansion.

\begin{lemma}[Sharp flat-endpoint witness]
\label{lem:flatendpoint}
Let $H\in C^2(0,a)$ be positive with $H,H'\to0$ as $x\downarrow0$. Every interval $(0,\epsilon)$, $\epsilon<a$, contains a point at which
\begin{equation}
 H''>0,\qquad HH''<(H')^2.
 \label{eq:flatwitness}
\end{equation}
The constant $1$ cannot be replaced uniformly by a smaller constant on this function class.
\end{lemma}
\begin{proof}
Suppose the strict inequality fails wherever $H''>0$. On a component $I$ of $\{H'>0\}$, the set $S=\{H''>0\}\cap I$ is nonempty, since $H'$ starts from zero at the left endpoint. It is open and relatively closed: at a putative boundary point inside $I$, the contrary inequality would give $0\ge(H')^2>0$. Thus $S=I$, and $H'$ is strictly increasing there. Every such component must reach $\epsilon$, so there is only one. Its left endpoint is zero, since $H'\le0$ before a positive left endpoint would contradict $H(0)=0$ and $H>0$. Therefore $H',H''>0$ throughout $(0,\epsilon)$ and $(\log H)''\ge0$. A convex function lies above its tangent at an interior point, which is incompatible with $\log H\to-\infty$ at a finite endpoint.

For sharpness take $H=e^{-1/x}$. Then $HH''/(H')^2=1-2x$ and $H''>0$ sufficiently near zero. Every constant smaller than $1$ fails eventually. This establishes sharpness within the stated function class.
\end{proof}

\begin{theorem}[Angular nonhyperbolicity arbitrarily near the center]
\label{thm:chargedangular}
Let a static, spherically symmetric solution of \eqref{eq:action} have bounded Kretschmann curvature and nonzero conserved charge $(P,Q)$. Suppose $h,W>0$ on a punctured center neighborhood and the responses are smooth at positive radii. Then every center neighborhood contains a point with a physical extraordinary angular characteristic satisfying $c_\perp^2<-1$.

The conclusion requires neither $A=B$, a curvature Taylor expansion, monotone curvature, nor $T\ne0$ throughout the neighborhood.
\end{theorem}

\begin{proof}
First let $P\ne0$. Theorem~\ref{thm:jointendpoint} and \eqref{eq:chargepotentialidentity} give
\begin{equation}
 H(R(r))\to0,\qquad
 H_R(R(r))=-\frac{r^4}{2P^2}V(r)\to0.
 \label{eq:Hflatendpoint}
\end{equation}
Since $H>0$, the curvature cannot equal its limiting value $R_0$ at a positive sufficiently small radius: the single-valued charge potential would then have both that positive value and its zero endpoint value. Hence it stays on one side, and we use $x=\pm(R-R_0)>0$.

The limits in \eqref{eq:Hflatendpoint} hold as curvature-function limits, not merely along a specially chosen sequence. For any small radial cutoff $r_1$, every curvature between $R_0$ and $R(r_1)$ is attained at a radius below $r_1$. The derivative identity and boundedness of $V$ give a uniform bound of order $r_1^4$ on that whole curvature interval. The modulus bound gives the corresponding limit of $H$ there. Thus Lemma~\ref{lem:flatendpoint} applies.

Choose its witness in the curvature interval attained below an arbitrary prescribed radial cutoff. At the corresponding positive radius, $H''>0$ and $HH''<H'^2$. Equation~\eqref{eq:Hspeed} is strictly below $-1$, since its last term is nonpositive. Also $L=-2b^2hH''\ne0$. If $T=0$ at this point, \eqref{eq:Heffectiveidentities} gives $H''=2h\kappa'^2>0$, so $\alpha\ne0$. Strict negativity of $M/L$ then gives $\beta_1\ne0$. The non-null negative root of the uneliminated matrix is therefore present; no exceptional $\alpha=0$ constraint branch is invoked.

For $P=0$, $Q\ne0$, set $H=1/h>0$. The same endpoint and derivative limits follow from $\mathcal C=Q^2H$. Here $b=0$, $T=2e^2h''$, and direct substitution into \eqref{eq:universalpolynomial} gives
\begin{equation}
 L=-\frac{2e^2H''}{H^3},\qquad
 c_\perp^2
 =1-\frac{2H'^2}{HH''}
 -\frac{2\theta'^2H^3}{H''}.
 \label{eq:electricHspeed}
\end{equation}
At a witness from the lemma, this is again strictly below $-1$. In fact $HH''<H'^2$ gives $h''>0$, hence $T>0$ there. This proves the purely electric case without a duality assumption.
\end{proof}

The theorem describes the classical equations at arbitrarily small positive radii. Section~\ref{Sec:BlackHoles} locates the same characteristic obstruction relative to black-hole horizons.

\begin{corollary}[Finite-order charged cores]
\label{cor:finiteorderinstability}
Suppose the positive effective potential has $H=\beta x^\nu(1+o(1))$, $\beta>0$, $\nu>1$, with controlled first and second differentiated asymptotics. Then, at every sufficiently small positive radius,
\begin{equation}
 c_\perp^2\le-\frac{\nu+1}{\nu-1}+o(1)<0.
 \label{eq:powerinstabilitybound}
\end{equation}
This includes $T=0$ points of the magnetic sector.
\end{corollary}

\begin{proof}
Here $H''>0$ and $H'^2/(HH'')\to\nu/(\nu-1)$. The last terms of \eqref{eq:Hspeed} and \eqref{eq:electricHspeed} are nonpositive. If $T=0$, the same argument used in Theorem~\ref{thm:chargedangular} establishes a physical non-null extraordinary root.
\end{proof}

For the cubic lift \eqref{eq:cubicliftlaw}, put $H=\beta x^3$, $\kappa=\eta x$. Then
\begin{equation}
 H''-2h\kappa'^2
 =\frac{2\beta x(3+2\eta^2x^2)}{1+\eta^2x^2}>0,
 \qquad
 \mathcal K=\frac{3\beta x^3}{3+2\eta^2x^2}>0,
\end{equation}
but
\begin{equation}
 c_\perp^2=-2-\frac43\eta^2x^2<0.
 \label{eq:cubicexactinstability}
\end{equation}
This is an angular gradient instability at ordinary finite-radius points despite positive $h$, positive $W$, and smooth central geometry. An infinitely flat response is also covered: for $H=e^{-1/x^2}$ and $\kappa=0$, one finds $c_\perp^2=(3x^2+2)/(3x^2-2)\to-1$.

The Legendre-degenerate set is not empty. An exact test is
\begin{equation}
 H=\beta x^2,\qquad
 \kappa=\frac12\left(\frac{x}{\ell_0}-\frac{\ell_0}{x}\right),
 \qquad \ell_0>0.
 \label{eq:Tzerolift}
\end{equation}
The responses are
\begin{equation}
 h=\frac{4\beta\ell_0^2x^4}{(x^2+\ell_0^2)^2},\qquad
 \theta=\theta_0-
 \frac{2\beta\ell_0x^3(x^2-\ell_0^2)}{(x^2+\ell_0^2)^2}.
 \label{eq:Tzeroresponses}
\end{equation}
Since $H''=2h\kappa'^2$, one has $T=0$ identically on the lifted background
\begin{equation}
 N=1-\frac{R_0}{12}r^2+cr^6,\qquad
 x=56cr^4,\qquad 224\beta cP^2=1.
 \label{eq:Tzerogeometry}
\end{equation}
With $\Lambda=0$, the lift proves all field equations; direct substitution gives $V=1$, $W=2$ and signed load $14cr^4$. At $W=2$, the auxiliary mixing coefficient $m$ vanishes. The constrained matrix therefore gives directly
\begin{equation}
 c_\perp^2=-\frac{(x^2+3\ell_0^2)^2}{4\ell_0^2x^2}<0.
 \label{eq:Tzeroexactinstability}
\end{equation}
This example tests precisely the branch on which a proof based only on division by $T$ would fail.

The $T=0$ example confirms the physical extraordinary root on a branch where the curvature Legendre map changes rank. The mechanism is curvature-response mixing in the coupled gravitational constraints, as distinct from the nonlinear dependence on electromagnetic invariants responsible for the angular instability in nonlinear electrodynamics~\cite{DeFeliceTsujikawa2025,DeFeliceTsujikawaScalar2025}. Together with the background endpoint theorem, this gives a dynamical restriction complementary to existence results for nonminimal charged configurations~\cite{Bokulic2026} and to general hyperbolicity criteria~\cite{Goulart2022}.

\subsection{Incompatible Cauchy normals}
A negative angular speed signals the obstruction in the static slicing. The full covector polynomial shows that the ordinary and extraordinary modes cannot share any hyperbolic time direction.

\begin{theorem}[No common Cauchy normal]
\label{thm:no-common-normal}
At a smooth point of a spherically symmetric background of \eqref{eq:action}, with $h>0$, $W>0$, $L\ne0$ and $c_\perp^2=M/L<0$, the ordinary and extraordinary physical characteristic quadratics have no common hyperbolicity direction. Consequently the complete physical characteristic system has no hyperbolic Cauchy normal there. In particular, a physically equivalent constrained formulation cannot be strongly hyperbolic on a metric-spacelike hypersurface through that point.
\end{theorem}
\begin{proof}
Up to nonzero factors the quadratics are
\begin{equation}
 Q_g(n)=-n_0^2+n_s^2+n_\perp^2,\qquad
 Q_e(n)=-n_0^2+n_s^2+c_\perp^2n_\perp^2.
 \label{eq:incompatible-quadratics}
\end{equation}
The hyperbolicity directions of $Q_g$, of signature $(-+++)$, satisfy $Q_g(n)<0$. Those of $Q_e$, of signature $(---+)$, satisfy $Q_e(n)>0$. But for every $Q_g$-timelike covector,
\begin{equation}
 Q_e(n)=Q_g(n)+(c_\perp^2-1)n_\perp^2<0.
\end{equation}
The cones are disjoint. A product polynomial is hyperbolic in a direction only when each factor is hyperbolic there. Since both factors belong to physical modes, adding gauge or constraint equations cannot remove this obstruction while preserving the physical solution space.
\end{proof}

\begin{figure}[t]
\centering
\includegraphics[width=.91\textwidth]{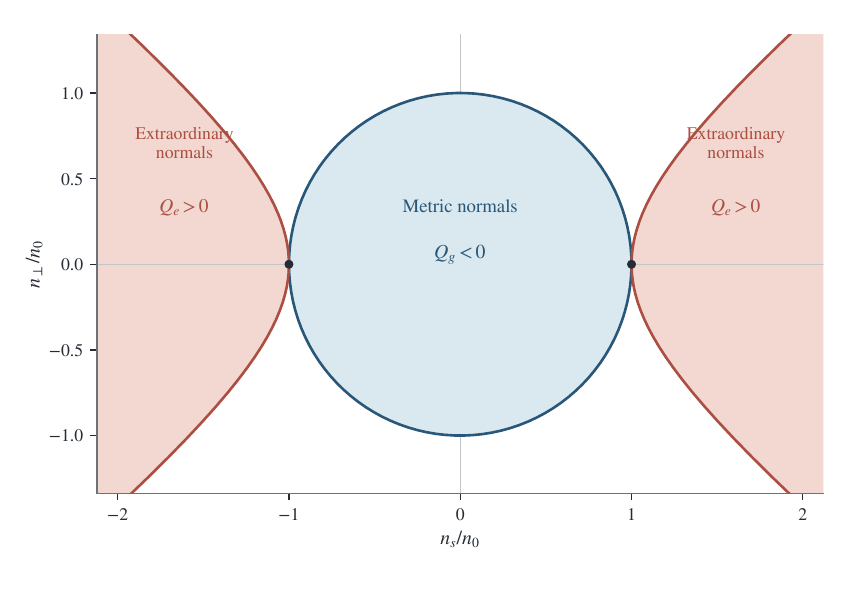}
\caption{Hyperbolicity directions in a cotangent section with $n_0=1$ and $c_\perp^2=-3/2$. The ordinary metric factor requires $Q_g<0$, whereas the extraordinary factor requires $Q_e>0$. Their open regions are disjoint, with boundaries meeting at radial null directions. The curves are sections of the spherical principal quadratics.}
\label{fig:incompatible-cones}
\end{figure}

The extraordinary quadratic is Lorentzian, with its distinguished hyperbolicity direction lying outside the ordinary timelike cone. Their disjoint open regions are shown in Fig.~\ref{fig:incompatible-cones}. The shared-cone criterion is familiar from electrodynamic hyperbolicity~\cite{Abalos2015}; the charged endpoint enforces its failure here, including on the $T=0$ branch.

At a fixed point the two nonreal angular frequencies of the full frozen linearized system satisfy
\begin{equation}
 \omega_\pm(k_\perp)=\pm i\sqrt{-c_\perp^2}\,|k_\perp|+O(1),
 \qquad |k_\perp|\longrightarrow\infty.
 \label{eq:frozen-growth}
\end{equation}
The $O(1)$ correction follows from the simple-root reduction in Appendix~\ref{App:PrincipalReduction}. For a spherical harmonic at radius $r_*$, $|k_\perp|=\sqrt{\ell(\ell+1)}/r_*$ in the local orthonormal frame. The nonreal root is the standard microlocal signal of a failure of Hadamard hyperbolicity~\cite{Metivier2005,Metivier2014,LernerNguyenTexier2018}. Equation~\eqref{eq:frozen-growth} is a local frozen-coefficient statement, not a calculation of a global quasinormal frequency. It proves a high-frequency obstruction for the classical equations as written. If those equations represent a truncated effective field theory, a mode can be interpreted within that theory only when a window of proper wavenumbers exists, $L_{\rm bg}^{-1}\ll |k_\perp|\ll\Lambda_{\rm EFT}$, in which the background varies slowly and omitted higher-derivative operators remain subleading. Neither the existence of that window nor the growth of modes below the cutoff follows from the principal polynomial alone. Conversely, a hypothetical ultraviolet completion does not make the untruncated classical principal system hyperbolic.

A simple nonreal physical root also persists under small changes of an existing smooth background family, at fixed slicing and tangential direction, provided the constraint rank remains regular. This fixed-slicing statement is more robust than the absence of every common normal. Indeed, for $a>0$ and any small $\delta>0$,
\begin{equation}
 Q_\delta(n)=-n_t^2+(1+\delta)n_s^2-a n_\perp^2
\end{equation}
shares hyperbolicity directions with $Q_g$ when $n_\perp=0$ and $1<n_t^2/n_s^2<1+\delta$. The spherical cones meet at radial null directions, allowing small perturbations of their principal coefficients to change the global cone-intersection property. The exact spherical structure is therefore essential to the slicing-independent theorem.

\section{Range of validity}\label{Sec:Validity}

\subsection{Positive mixing of Abelian fields}
The obstruction is not peculiar to a single Maxwell field. Consider finitely many Abelian field strengths with curvature-dependent kinetic and odd-response mixing, described by the fixed action
\begin{equation}
 S=\frac1{16\pi}\int\sqrt{-g}\,
 \left[R-2\Lambda-K_{IJ}(R)F^I_{ab}F^{Jab}
 -\Theta_{IJ}(R)F^I_{ab}{}^\star F^{Jab}\right]\,\dd^4x,
 \label{eq:multi-action}
\end{equation}
where $K$ and $\Theta$ are real symmetric matrices. We assume $K>0$ at every sufficiently small positive radius; no uniform lower eigenvalue bound is assumed. The conserved charges are vectors,
\begin{equation}
 \mathbf b=\frac{\mathbf P}{r^2},\qquad
 \mathbf Q=r^2K\mathbf e+\Theta\mathbf P.
\end{equation}
For fixed $\mathbf P,\mathbf Q$ define
\begin{equation}
 \mathcal C=\mathbf P^TK\mathbf P+
 (\mathbf Q-\Theta\mathbf P)^TK^{-1}(\mathbf Q-\Theta\mathbf P).
 \label{eq:multi-charge}
\end{equation}
This is positive for a nonzero combined charge vector, even when no constant field basis diagonalizes the responses throughout the solution.

\begin{theorem}[Arbitrary positive Abelian mixing]
\label{thm:multi}
Let a static, spherically symmetric solution of \eqref{eq:multi-action} have bounded Kretschmann curvature and $(\mathbf P,\mathbf Q)\ne(0,0)$. Assume smooth response matrices and fields at positive radii, $K>0$, and $W=f_R>0$ throughout a sufficiently small punctured neighborhood. Every center neighborhood contains a positive-radius physical extraordinary angular root with $c_\perp^2<-1$. At that point the full physical system has no common hyperbolic normal. The conclusion does not require $T=f_{RR}$ to remain nonzero throughout the neighborhood.
\end{theorem}

\begin{proof}
The algebraic odd-response stress cancels as in the single-field theory. The radial-minus-temporal equation and trace are unchanged, with
\begin{equation}
 V=-K_{IJ,R}U^{IJ}-\Theta_{IJ,R}G^{IJ},\qquad W=1+V.
\end{equation}
The bounded-source argument therefore gives bounded $V$ and $\mathcal E=\mathbf b^TK\mathbf b+\mathbf e^TK\mathbf e=O(1)$. Differentiating the fixed-charge expression gives
\begin{equation}
 \mathcal C=r^4\mathcal E,\qquad
 \mathcal C_R=-\frac{r^4}{2}V.
 \label{eq:multi-derivative}
\end{equation}
In particular, both $\mathcal C$ and $\mathcal C_R$ tend to zero. If bounded continuous $R(r)$ had two distinct accumulation values, every intermediate value $c$ would occur along radii tending to zero, forcing the fixed positive number $\mathcal C(c)$ to vanish. Thus $R\to R_0$. Positivity excludes attaining $R_0$ at a sufficiently small positive radius. The intermediate-value argument used above upgrades the limits to the entire one-sided attained curvature interval.

Put
\begin{align}
 \boldsymbol\alpha&=K_R\mathbf e+\Theta_R\mathbf b,&
 \boldsymbol\beta&=K_R\mathbf b-\Theta_R\mathbf e,\notag\\
 \mathcal A&=\boldsymbol\alpha^TK^{-1}\boldsymbol\alpha,&
 \mathcal B&=\boldsymbol\beta^TK^{-1}\boldsymbol\beta.
\end{align}
The uneliminated auxiliary action contains one curvature multiplier. For a non-null covector $q$, let
\begin{equation}
 J_I^{ab}=K_{IJ,R}F^{Jab}+\Theta_{IJ,R}{}^\star F^{Jab},
 \qquad j_I^b=q_aJ_I^{ab}.
\end{equation}
Antisymmetry implies $q_bj_I^b=0$. For a non-null covector, the Lorenz-gauge Maxwell equation gives
\begin{equation}
 a_I^b=\frac{i z}{q^2}(K^{-1})_{IJ}j_J^b,\qquad z=\delta\chi.
\end{equation}
The retained curvature constraint then yields
\begin{equation}
 Tq^2+4(K^{-1})_{IJ}j_I\cdot j_J=0.
 \label{eq:multi-covector}
\end{equation}
The scalar and harmonic-gauge metric feedback have the same lower orders on this non-null branch as in the covariant calculation above. No division by $T$ occurs. Radial symmetry evaluates the contraction as $\mathcal A(\omega^2-k_r^2)+\mathcal Bk_\perp^2$, so the physical factor is
\begin{equation}
 \mathcal D(\omega^2-k_r^2)-\mathcal N k_\perp^2,
 \quad \mathcal D=T-4\mathcal A,\quad \mathcal N=T+4\mathcal B.
 \label{eq:multi-cone}
\end{equation}
On a nondegenerate Legendre branch the rank-one correction leaves the other $2n-1$ photon polarizations luminal, where $n$ is the number of gauge fields. The non-null extraordinary mode is also defined by the uneliminated system at $T=0$.

Differentiating the flux expression a second time gives the identity
\begin{equation}
 \mathcal D=-\frac{2\mathcal C_{RR}}{r^4},\qquad
 \frac{\mathcal C_R}{r^4}
 =\mathbf b^T\boldsymbol\beta-\mathbf e^T\boldsymbol\alpha.
 \label{eq:multi-Hessian}
\end{equation}
Cauchy--Schwarz in the positive $K$ norm takes a particularly useful form:
\begin{align}
 \mathcal A+\mathcal B-\frac{\mathcal C_R^2}{r^4\mathcal C}
 ={}&\left\|K^{-1/2}\left(\boldsymbol\beta-\frac{\mathcal C_R}{\mathcal C}K\mathbf b\right)\right\|^2\notag\\
 &+\left\|K^{-1/2}\left(\boldsymbol\alpha+\frac{\mathcal C_R}{\mathcal C}K\mathbf e\right)\right\|^2\ge0.
 \label{eq:multi-CS}
\end{align}
At $\mathcal C_{RR}>0$ this gives
\begin{equation}
 c_\perp^2=\frac{\mathcal N}{\mathcal D}
 =1-\frac{2r^4(\mathcal A+\mathcal B)}{\mathcal C_{RR}}
 \le1-\frac{2\mathcal C_R^2}{\mathcal C\mathcal C_{RR}}.
 \label{eq:multi-bound}
\end{equation}
Lemma~\ref{lem:flatendpoint} supplies an attained witness with $0<\mathcal C\mathcal C_{RR}<\mathcal C_R^2$. Hence $\mathcal D<0$, $\mathcal N>0$, and $c_\perp^2<-1$. If $T=0$ there, then $\mathcal A>0$ and $\mathcal B>0$; the displayed non-null Maxwell amplitude has nonzero field strength. Thus the root survives the auxiliary constraint. The two incompatible normal inequalities in Theorem~\ref{thm:no-common-normal} apply unchanged to \eqref{eq:multi-cone}.
\end{proof}

The same argument covers electric, magnetic and mixed charge vectors. The additional gauge-field directions enter the extraordinary speed through the nonnegative remainder in \eqref{eq:multi-CS}. The theorem assumes positive-definite $K$; degenerate or indefinite response matrices and non-Abelian fields require a different analysis~\cite{BalakinYangMills2016a,BalakinYangMills2016b}.

\subsection{Beyond the Einstein--Hilbert action}
The gravitational curvature Hessian changes the characteristic problem as well. For a general scalar-curvature Lagrangian~\cite{SotiriouFaraoni2010,DeFeliceTsujikawaReview2010}, we construct an exact local example with
\begin{equation}
 S=\frac1{16\pi}\int\sqrt{-g}\,[\Phi(R)-h(R)U]\,\dd^4x.
 \label{eq:Phi-counter-action}
\end{equation}
Here $W=\Phi_R-h_RU$ and $T=\Phi_{RR}-h_{RR}U$. The auxiliary calculation changes the extraordinary coefficient to
\begin{equation}
 \mathcal D=\Phi_{RR}-\frac{2\mathcal C_{RR}}{r^4}.
 \label{eq:Phi-principal-shift}
\end{equation}
The additional gravitational Hessian can therefore compete with the curvature response in the extraordinary coefficient. Our example has positive principal kinetic and gradient coefficients and a subluminal extraordinary photon cone throughout a punctured center neighborhood. Positive small-scale principal signs have also been obtained in other scalar--electromagnetic models~\cite{DeFeliceTsujikawaScalar2025}.

To see that this change can genuinely restore a common normal, we construct one fixed analytic action with a magnetically charged, analytically regular central geometry. With $P=1$ and $u=r^2$, the solution has
\begin{equation}
 W=1+\frac u2,\qquad B=1-u+\frac{13}{20}u^2-\frac{17}{32}u^3,
 \qquad A=(1+u)^{1/2}B.
 \label{eq:Phi-example-brief}
\end{equation}
Its curvature obeys $R(0)=9$ and $R_u(0)=-9$, so the rational functions $h(R(u))$ and $\Phi(R(u))$ in Appendix~\ref{app:phi-example} define a single-valued analytic action near $R=9$. The full metric and Maxwell equations, including the constitutive identity, hold exactly. On $0<u\le10^{-3}$, we find
\begin{equation}
 h>0,\qquad W>0,\qquad T<0,\qquad 0<c_\perp^2<1,\qquad c_\perp^2\longrightarrow\frac35.
 \label{eq:Phi-example-summary}
\end{equation}
The other propagating cones are luminal, so static time defines a common hyperbolicity direction throughout the punctured neighborhood. The metric extends analytically across the center, whereas the charged magnetic two-form retains its nonzero flux. The photon kinetic coefficient $h$ tends to zero there. This exact local solution shows why the Einstein--Hilbert assumption is essential to the hyperbolicity obstruction.

\section{Black holes and global constraints}\label{Sec:BlackHoles}

We now ask how the local characteristic obstruction appears in black-hole geometries. Two global identities constrain the connection between a charged center and an asymptotic end. We then use a fixed-action family built from a known Sert solution to locate the extraordinary cone relative to the horizons.

\subsection{One-function rigidity and weak-curvature response}
For the single-response theory, suppose the geometry has a connected ingoing chart
\begin{equation}
 ds^2=-N(r)\,dv^2+2\,dv\,dr+r^2d\Omega^2.
 \label{eq:fa-ingoing}
\end{equation}
The one-function condition is a restriction on the geometry in areal radius.

\begin{proposition}[One-function rigidity]
\label{prop:fa-rigidity}
Suppose the center in Eq.~\eqref{eq:fa-ingoing} has bounded Kretschmann curvature and the composite response $W=1-h_RU$ is $C^2$. Then $W$ is constant throughout the connected chart. If $R$ is nonconstant, necessarily
\begin{equation}
 \Lambda=0,\qquad W=2,\qquad -h_RU=1.
 \label{eq:fa-Wtwo}
\end{equation}
\end{proposition}

\begin{proof}
The covariant $rr$ equation in the ingoing chart has $g_{rr}=R_{rr}=T_{rr}=0$ and $\nabla_r\nabla_rW=W''$. It therefore gives $W''=0$, including at a multiple horizon. The central estimate $W'=O(r)$ fixes the affine slope to zero. The trace reduces to $(2-W)R=4\Lambda$; a nonconstant $R$ implies Eq.~\eqref{eq:fa-Wtwo}.
\end{proof}

If $R$ is constant, its value is attained at every positive radius and $h$ is finite there. Bounded weighted load then excludes electric displacement charge and requires $h=0$ for nonzero magnetic charge. This is the screened exception, and it cannot approach a nonzero response at another radius of the same constant-curvature solution.

On the purely magnetic nonzero-response branch, defined by $Q=0$ and $P\ne0$, the relation following from Eq.~\eqref{eq:fa-Wtwo} is
\begin{equation}
 h_R(R(r))=-\frac{r^4}{2P^2}.
 \label{eq:fa-hR}
\end{equation}
On the purely electric nonzero-response branch, defined by $P=0$ and $Q\ne0$, the corresponding relation is instead
\begin{equation}
 j_R(R(r))=-\frac{r^4}{2Q^2},\qquad j=\frac1h.
 \label{eq:fa-jR}
\end{equation}
These identities concern the two separate one-charge branches and make no claim for a dyonic branch with $PQ\ne0$. On either nonzero-response pure branch, two distinct radii therefore cannot have the same scalar curvature: the derivative of the corresponding single-valued response would take two different values at that curvature. In particular, a curvature-turning reconstruction cannot define a global $C^1$ action on either branch. Screened intervals are excluded from this argument because their electric field need not vanish.

For comparison, the familiar one-function magnetic reconstruction is
\begin{equation}
 R=-N''-\frac{4N'}r-\frac{2(N-1)}{r^2},\qquad
 h_m=\frac{r^4}{2P^2}
       \left[N''-\frac{2(N-1)}{r^2}\right].
 \label{eq:fa-reconstruction}
\end{equation}
Its chain-rule identity
\begin{equation}
 \frac{dh_m}{dr}=-\frac{r^4}{2P^2}\frac{dR}{dr}
\end{equation}
holds for every sufficiently differentiable $N$. A radial construction still has to obey Eq.~\eqref{eq:fa-hR} at repeated curvature values; the chain rule alone does not ensure a global response.

\begin{corollary}[Weak-curvature response obstruction]
\label{cor:fa-cusp}
A one-function solution connecting a bounded-curvature charged center to an asymptotic end $r\to\infty$ with finite nonzero response cannot have bounded $h_R$ at that end. On a magnetic exterior, $h_R=-r^4/(2P^2)$; on an electric exterior, $h_R=r^4h^2/(2Q^2)$.
\end{corollary}

If, more specifically, $R=\kappa r^{-n}(1+o(1))$ and $R'=-n\kappa r^{-n-1}(1+o(1))$, with $\kappa>0$, the magnetic response has a finite limit only if $n>4$, and then
\begin{equation}
 h_\infty-h(r)=
 \frac{n\kappa}{2P^2(n-4)}r^{4-n}(1+o(1)).
 \label{eq:fa-cusp-rate}
\end{equation}
Its curvature cusp exponent is $1-4/n$. The same formula applies to $1/h$ in the electric sector. The derivative asymptotic is an explicit hypothesis here; it is not inferred by differentiating a little-$o$ term.

Write $A=C^2B$, with $C>0$, and retain the regular null chart through finite-radius horizons. The radial-temporal equation is
\begin{equation}
 W''-\frac{C'}C W'-\frac{2C'}{rC}W=0.
 \label{eq:fa-general-W}
\end{equation}

\subsection{A trapped-region trace identity}
The trace equation also gives a global restriction that does not impose $A=B$. Set $\Lambda=0$ and use the regular null form
\begin{equation}
 ds^2=-C^2B\,dv^2+2C\,dv\,dr+r^2d\Omega^2,\qquad C>0.
\end{equation}
Suppose $r_-<r_+$ are regular simple horizons, $B<0$ between them, and $B>0$ on a connected exterior. For the classical integration below, take $C,B\in C^1$ and $W\in C^2$ in the regular ingoing chart, with $R$ bounded and continuous. We also require $C$ to be finite and nonzero through both horizons. These assumptions make the trace equation classical across the horizons and ensure that $W$ and $W'$ are bounded there. With $Y=W-2$, the exact trace reads
\begin{equation}
 3(Cr^2B Y')'+Cr^2RY=0.
 \label{eq:trapped-trace-ode}
\end{equation}
Multiplying by $Y$ and integrating gives
\begin{equation}
 \int_{r_-}^{r_+}Cr^2
 \left[3(-B)(W')^2+R(W-2)^2\right]dr=0.
 \label{eq:trapped-trace-integral}
\end{equation}
The boundary term vanishes at the regular horizons. The sign of $B$ is essential: the derivative term is nonnegative in the trapped region.

\begin{theorem}[Trapped curvature and asymptotic response]
\label{thm:trapped-curvature}
Under these hypotheses, if $R\ge0$ between the horizons and is positive somewhere, then $W=2$ throughout the trapped interval and its connected exterior. Such a solution cannot have $W\to1$ at its far end.

If also $h(0)=1$, $\theta(0)=0$, their curvature derivatives are finite, and
\begin{equation}
 \mathcal C_R(0)=(P^2-Q^2)h_R(0)-2PQ\theta_R(0)\ne0,
 \label{eq:far-charge-derivative}
\end{equation}
then any solution with $W\to1$ must have $R<0$ somewhere between its horizons.
\end{theorem}

\begin{proof}
When $R\ge0$, both terms of \eqref{eq:trapped-trace-integral} are nonnegative. Thus $W$ is constant in the trapped interval, and any point with $R>0$ fixes this constant to two.

Extension across the outer horizon needs a separate argument. Put $p=Cr^2B$ and $q=Cr^2R$. Simplicity of the horizon gives $|p(r)|\ge c(r-r_+)$ just outside it, while $q$ is bounded. Since $Y(r_+)=0$ and $pY'(r_+)=0$,
\begin{equation}
 p(r)Y'(r)=-\frac13\int_{r_+}^{r}q(s)Y(s)ds.
\end{equation}
For $M_\epsilon=\sup_{[r_+,r_++\epsilon]}|Y|$, this gives $|Y'|\le K M_\epsilon$ and hence $M_\epsilon\le K\epsilon M_\epsilon$. Taking $\epsilon$ small proves $Y=0$ across the horizon. Ordinary linear uniqueness then extends this identity through the connected exterior.

The remaining possibility is $R\equiv0$ on the trapped interval. The integral still gives constant $W$, but the charge identity is now $W-1=-2\mathcal C_R(0)/r^4$. Constancy on a nonempty radial interval therefore requires $\mathcal C_R(0)=0$ and $W=1$, contrary to \eqref{eq:far-charge-derivative}. This proves the final assertion.
\end{proof}

The zero-curvature exception is necessary: Reissner--Nordstrom has $R=0$, $W=1$ and $\mathcal C_R(0)=0$ between its horizons. The known Sert lift provides the other exact check, with $R>0$ and $W=2$; it fails the normal far-end condition rather than the theorem.

For example, a fixed purely magnetic cubic response $h(R)=(1-R/R_c)^3$ with $\theta=0$ has $\mathcal C_R(0)=P^2h_R(0)=-3P^2/R_c\ne0$. A two-horizon completion joining $R_c>0$ to an asymptotic end with $W\to1$ must therefore develop negative scalar curvature in its trapped region. Such a completion would require a nonmonotone curvature profile; its existence remains an open global question.

\subsection{A fixed-action near-extremal family}
Fix $R_c,H_\infty>0$ and $\Lambda=0$. On the positive-curvature interval $0<R<R_c$, take the seed response
\begin{equation}
 H(R)=H_\infty
 \left[1-\left(\frac R{R_c}\right)^{3/7}\right]^{7/3},
 \qquad 0<R<R_c.
 \label{eq:fa-Hglobal}
\end{equation}
The first Sert family~\cite{Sert2016} has
\begin{align}
 N&=1-\frac{2m}{r}\left[1-(1+a^3r^3)^{-1/3}\right],
 &R&=\frac{8ma^3}{(1+a^3r^3)^{7/3}},\label{eq:fa-Sert}\\
 H(r)&=\frac{4ma^6r^7}{P^2(1+a^3r^3)^{7/3}}.
\end{align}
It belongs to the same fixed action for different charges exactly when
\begin{equation}
 R_c=8ma^3,\qquad H_\infty=\frac{4m}{aP^2},
 \quad\text{or}\quad
 a^4P^2=\frac{R_c}{2H_\infty},\qquad m=\frac{R_c}{8a^3}.
 \label{eq:fa-parameters}
\end{equation}
Thus $P$ is free, while
\begin{equation}
 m(P)=\frac{(2H_\infty)^{3/4}R_c^{1/4}}8|P|^{3/2}.
\end{equation}
The response parameters are fixed, while $m$ and $a$ vary with the charge. Moreover, $R'=-7R_ca^3r^2/(1+a^3r^3)^{10/3}<0$ for all $r>0$, so the action remains single-valued along the entire solution.

\begin{figure}[t]
\centering
\includegraphics[width=.99\textwidth]{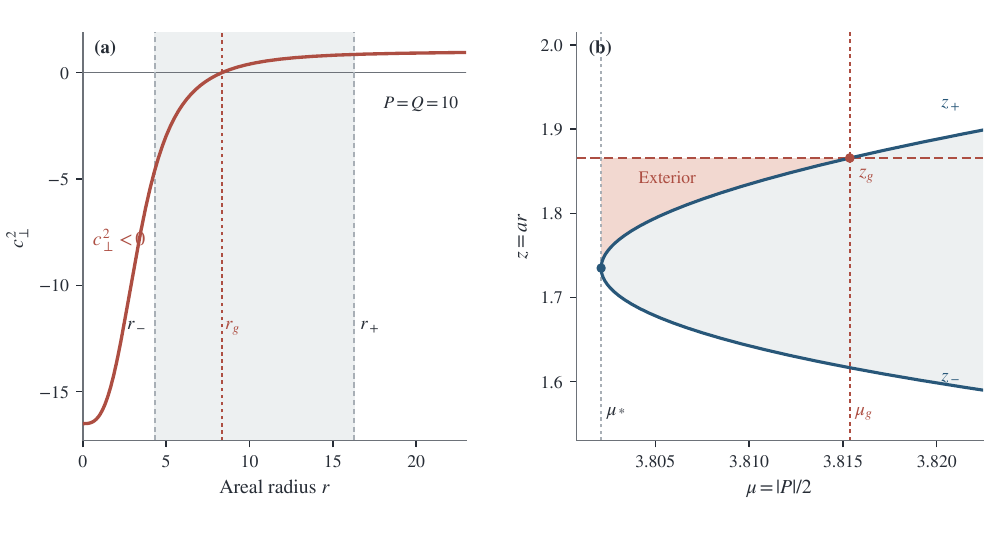}
\caption{The extraordinary characteristic in the fixed action with $R_c=1$, $H_\infty=2$ and $\theta_0=\eta=1$. Left: squared angular speed for $P=Q=10$, with the two horizons and its zero $r_g$. Right: horizon positions and the cone-change radius in dimensionless coordinates. The interval $\mu_*<\mu<\mu_g$ has a negative-cone region outside the outer horizon.}
\label{fig:blackholes}
\end{figure}

Choose fixed constants $\theta_0\ne0,\eta$, set $H_\infty=1+\theta_0^2$, and apply the two-response lift:
\begin{equation}
 \kappa(R)=\theta_0+\eta R/R_c,\qquad
 h=\frac{H}{1+\kappa^2},\qquad
 \theta=\theta_0-\kappa h.
 \label{eq:fa-pair}
\end{equation}
The fields and charges are
\begin{equation}
 b=\frac P{r^2},\qquad e=\frac{\kappa P}{r^2},\qquad
 Q=\theta_0P.
 \label{eq:fa-fields}
\end{equation}
Using $U=2(b^2-e^2)$ and $G=-4eb$, direct substitution gives
\begin{equation}
 h(e^2+b^2)=\frac{HP^2}{r^4},\qquad
 -h_RU-\theta_RG=-\frac{2P^2H_R}{r^4}=1,
 \qquad r^2he+\theta P=\theta_0P.
 \label{eq:fa-liftcheck}
\end{equation}
The temporal, radial, angular and trace equations coincide with those of the exact Sert seed; the flux identity verifies the Maxwell equations. For all charges in the family the two response functions define the same action on the attained interval $0<R<R_c$. The fields carry both electric and magnetic flux, and the resulting horizon and characteristic structure is shown in Fig.~\ref{fig:blackholes}. Ref.~\cite{Bokulic2026} also notes that the magnetic Sert solution evades its regularity no-go condition because the curvature derivative of the response vanishes at the center. Here the same seed serves a different purpose: the two-response lift preserves the background solution while allowing the characteristic cone of the coupled perturbations to be examined.

For example, $\theta_0=\eta=1$ makes $\kappa$ nonzero throughout. The ratio $H/\kappa$ runs from zero at the center to two at infinity. It fails the constant-transformation criterion for removing the electric field, so this example is not merely a constant topological shift of a pure magnetic configuration.

The central expansion and curvature are
\begin{equation}
 N=1-\frac{R_cr^2}{12}+\frac{R_ca^3r^5}{18}+O(r^8),
 \qquad K\longrightarrow\frac{R_c^2}{6}.
 \label{eq:fa-center}
\end{equation}
All orthonormal Riemann components are bounded. The Cartesian metric is $C^4$ but not $C^5$, owing to the nonzero $r^5$ term. Nonzero flux still precludes a smooth source-free field extension over the filled center.

Put $z=ar$, $\mu=2ma=|P|\sqrt{2H_\infty R_c}/4$, and write
\begin{equation}
 N=1-\mu f(z),\qquad
 f(z)=\frac{1-(1+z^3)^{-1/3}}z.
\end{equation}
The positive function $f$ vanishes at both endpoints. With $s=(1+z^3)^{1/3}>1$, its unique stationary point satisfies
\begin{equation}
 s^3-s^2-s-1=0.
\end{equation}
Indeed the derivative of this polynomial is $(3s+1)(s-1)>0$ for $s>1$, and the logarithmic derivative of $f$ has the opposite sign to the polynomial. If $s_*$ denotes its root, define
\begin{equation}
 z_*=(s_*^3-1)^{1/3},\qquad
 \mu_* =\frac{s_*z_*}{s_*-1}=3.8021155\ldots.
 \label{eq:fa-horizon-threshold}
\end{equation}
There are no horizons for $\mu<\mu_*$, one double horizon for $\mu=\mu_*$, and exactly two simple horizons for $\mu>\mu_*$. Both metric and electromagnetic field extend across the finite-radius horizons in Eq.~\eqref{eq:fa-ingoing}, where $F=-e\,dv\wedge dr+P\sin\vartheta\,d\vartheta\wedge d\varphi$.

For the fixed dimensionless theory
\begin{equation}
 R_c=1,\quad H_\infty=2,\quad\theta_0=\eta=1,
 \label{eq:fa-concrete-theory}
\end{equation}
the choice $P=Q=10$ gives
\begin{equation}
 a\simeq0.223607,\qquad m\simeq11.180340,\qquad\mu=5,
\end{equation}
and
\begin{equation}
 r_-\simeq4.328870,\qquad r_+\simeq16.248426.
\end{equation}

At infinity,
\begin{equation}
 N=1-\frac{2m}r+\frac{2m}{ar^2}+O(r^{-5}),\qquad
 R\sim\frac{R_c}{a^7r^7}.
\end{equation}
The geometric ADM mass is $m$, and $h\to1$, $\theta\to0$. The limiting response values are Maxwell-like, while their curvature derivatives behave differently. With $y=R/R_c$,
\begin{align}
 h&=1-\frac73 y^{3/7}+o(y^{3/7}),&
 h_R&\sim-\frac1{R_c}y^{-4/7},\\
 \theta&=\frac73\theta_0y^{3/7}+o(y^{3/7}),&
 \theta_R&\sim\frac{\theta_0}{R_c}y^{-4/7}.
 \label{eq:fa-weak-cusp}
\end{align}
Thus the response functions approach Maxwell values but have a curvature cusp at $R=0$, while $W=2$ on these solutions. Their finite limiting values do not imply a smooth weak-curvature response: both $h_R$ and $\theta_R$ diverge. The family has $W=2$, rather than the asymptotic condition $W\to1$ in Theorem~\ref{thm:trapped-curvature}; it also lies outside that theorem's second conclusion, which assumes finite $h_R(0)$ and $\theta_R(0)$. There is therefore no conflict with the trapped-region result. The charge term in the metric is $(P^2+Q^2)/(2r^2)$ at the bare gravitational normalization. The family realizes the cusp in Corollary~\ref{cor:fa-cusp} rather than a smooth weak-curvature Einstein--Maxwell limit.

In the concrete theory~\eqref{eq:fa-concrete-theory}, let
\begin{equation}
 D_H=H_{RR}-2h\kappa_R^2,\qquad
 \mathcal K=\frac{hH_{RR}}{D_H},\qquad
 \mathcal G_\perp=h-
 \frac{2(H_R-h\kappa\kappa_R)^2}{D_H}.
 \label{eq:fa-cone-coeffs}
\end{equation}
These are the transverse extraordinary-photon coefficients, with $\kappa_R=1$. The Legendre coefficient is $T=-2P^2D_H/r^4$. It is nonzero at every positive radius. To see this, put $u=R^{3/7}\in(0,1)$ and note
\begin{equation}
 \frac{H_{RR}}h
 =\frac47[1+(1+R)^2]u^{-11/3}(1-u)^{-2}
 \ge\frac8{7u^2(1-u)^2}\ge\frac{128}7>2.
\end{equation}
Thus $D_H>0$ and $\mathcal K>0$. Nevertheless, setting $t=R^{1/7}$ in Eq.~\eqref{eq:fa-cone-coeffs} gives the exact result
\begin{equation}
 c_\perp^2=\frac{\mathcal G_\perp}{\mathcal K}
 =1-7t^3-7t^7-\frac72t^{11}.
 \label{eq:fa-global-cone}
\end{equation}
It decreases strictly with $R$ and has one zero, $R_g\simeq0.009100206$. For $P=10$ this occurs at $r_g\simeq8.343422$, between the two horizons. The speed approaches one at infinity and $-33/2$ at the center. In particular, the inner static region has a negative extraordinary angular speed despite positive $h$, $W$ and $\mathcal K$.

The cone-change radius has a fixed dimensionless position throughout this fixed-theory family. Let $t_g\in(0,1)$ be the unique root of
\begin{equation}
 1-7t_g^3-7t_g^7-\tfrac72t_g^{11}=0,
 \qquad
 z_g=(t_g^{-3}-1)^{1/3},\qquad \mu_g=\frac{z_g}{1-t_g}.
 \label{eq:cone-horizon-threshold}
\end{equation}
Since $s_*<19/10$ and hence $s_*^3<7$, the cone polynomial is negative at $t=1/s_*$. Its strict monotonicity implies $z_g>z_*$, and the decrease of $f(z)$ for $z>z_*$ implies $\mu_g>\mu_*$. Numerically,
\begin{equation}
 z_g\simeq1.865646,\qquad \mu_g\simeq3.8153644.
\end{equation}
For $\mu_*<\mu<\mu_g$, the two horizons satisfy $z_-<z_*<z_+<z_g$. Thus the static exterior annulus $r_+<r<r_g$ has a negative angular cone. At $\mu=\mu_g$ the sign change meets the outer horizon; for $\mu>\mu_g$ it lies between the two horizons. In the chosen fixed theory, the exterior-annulus band is
\begin{equation}
 7.60423<|P|<7.63073.
\end{equation}
Throughout this near-extremal interval the exterior annulus carries the incompatible extraordinary characteristic, while the response law and its weak-curvature cusp remain fixed across the family.

A related geometric no-go theorem for regular black holes in scalar--tensor theories was recently obtained by Peng et al.~\cite{Peng2026}; there the obstruction is tied to the odd-parity tensor characteristic at an interior turning point of the trapped region. The mechanism here is different: the relevant mode is the extraordinary electromagnetic polarization generated by the curvature-dependent Maxwell response, and the exact fixed-action family places its negative cone in a finite annulus outside the outer horizon near extremality.

\section{Discussion}\label{Sec:Discussion}
A charged spherical center can be mild in its curvature and yet fail a more basic dynamical test. In the Einstein--electromagnetic theory studied here, the background equations force a positive charge potential $\mathcal C$ and its curvature derivative to vanish at the central endpoint. This flattening controls more than the local electromagnetic stress: it produces a physical extraordinary characteristic with $c_\perp^2<-1$ at radii arbitrarily close to the center. The ordinary photon and graviton cones remain metric cones, but the extraordinary quadratic has no hyperbolicity direction in common with them. The obstruction belongs to the coupled classical initial-value problem and cannot be removed by choosing another spacelike slicing.

This result complements restrictions on the existence of nonminimally coupled charged backgrounds and on regular solutions in nonlinear electrodynamics~\cite{Bokulic2022,Bokulic2026,BokulicNLE2026}. The recent trace-based analysis of Ref.~\cite{Bokulic2026} treats several specified curvature-coupling sectors and identifies simultaneous dependence on both electromagnetic invariants as a further nontrivial extension. The present action realizes a separable two-response version of that extension and asks a different question: once a charged background satisfies the field equations, do its physical perturbations admit a common Cauchy normal? The distinguishing ingredient is the curvature derivative of the Maxwell response, which couples the extraordinary electromagnetic polarization to the gravitational curvature constraint. The full constraint reduction retains this physical root even when $f_{RR}=0$, and positive-definite mixing of Abelian fields strengthens the same endpoint inequality. The resulting failure of a common hyperbolicity direction connects these charged-center bounds with characteristic-cone methods in electrodynamics and gravitational evolution~\cite{Abalos2015,Goulart2022,ReallTanahashiWay2014,PapalloReall2017,KovacsReall2020}.

The exact solutions illustrate how local and global aspects fit together. A center may have bounded curvature even though its conserved flux prevents a smooth source-free electromagnetic extension over the filled ball. In the fixed-action family based on Sert's geometry~\cite{Sert2016}, a finite interval of near-extremal charges places the negative extraordinary cone outside the outer horizon. The same family has a cusp in its response functions at vanishing curvature, consistent with the independent global restrictions derived from the trace equation. These features locate the hyperbolicity obstruction in a concrete geometry while showing the importance of the response law away from the center.

The analytic $\Phi(R)$ construction reveals the role of the gravitational action: an additional curvature Hessian can yield a real, subluminal extraordinary photon cone near a charged center. More general curvature-tensor and vector--tensor couplings have their own principal systems~\cite{Verbin2022,Chen2026,Colleaux2025,BuenoCanoHennigar2025,FrolovKoekSotoZelnikov2025}. There is also a scale question when curvature--photon interactions are treated as effective corrections~\cite{DrummondHathrell1980,RussoTownsend2024,RussoTownsend2026}: the classical frozen-coefficient growth probes arbitrarily large wave numbers, whereas physical predictions from a truncated theory require a window below its cutoff. Asymptotic photon propagation provides a complementary probe~\cite{CarballoRubio2025}. Finding a charged configuration with bounded central curvature, compatible physical characteristic cones and a controlled weak-curvature limit is therefore a sharper goal than background regularity alone.

\appendix

\section{Physical non-null characteristics and the constraint reduction}\label{App:PrincipalReduction}

We verify that the extraordinary non-null root of Sect.~\ref{Sec:Hyperbolicity} belongs to the physical constrained system, including at $T=f_{RR}=0$. The calculation has three ingredients: elimination of the invertible non-null wave blocks, identification of the remaining one-dimensional Maxwell--curvature kernel, and a check of the gauge subsidiary equations. It also establishes persistence of every simple negative-angular root when frozen lower-order terms are restored.

For $W>0$, the uneliminated auxiliary action \eqref{eq:uneliminatedaction} takes the Einstein-frame form
\begin{equation}
 S_{\rm aux}=\frac1{16\pi}\int\!\sqrt{-\widetilde g}\left[
 \widetilde R-\frac{3}{2W^2}(\widetilde\nabla W)^2
 -\frac{(W-1)\chi+2\Lambda}{W^2}
 -h(\chi)\widetilde U-\theta(\chi)\widetilde G\right]\dd^4x.
 \label{eq:app-Einstein-frame}
\end{equation}
The conformal transformation is invertible because $W>0$. Varying the original auxiliary action with respect to $W$ enforces $\chi=R$, while varying $\chi$ gives $W=f_\chi$. Both statements remain valid when $f_{\chi\chi}=T=0$. The auxiliary system therefore has the same physical solutions as the original theory even where the usual curvature Legendre inversion fails.

Let $p$ be a smooth background point and take a Fourier covector $q_a=\lambda\widehat q_a$ with $\widehat q^2\ne0$ and $\lambda\to\infty$. Use Einstein harmonic gauge for $\delta\widetilde g_{ab}$ and Lorenz gauge for $a_b=\delta A_b$. In trace-reversed metric variables, the diagonal symbols of the tensor, propagating scalar and Maxwell blocks are, up to nonzero normalizations,
\begin{equation}
 \sigma_2(E_{\rm grav})\sim\widetilde q^2\overline{\delta\widetilde g}_{ab},\quad
 \sigma_2(E_W)\sim\frac{3}{W^2}\widetilde q^2\delta W,\quad
 \sigma_2(E_A)\sim h\widetilde q^2a^b.
 \label{eq:app-diagonal-blocks}
\end{equation}
All three blocks are invertible on the non-null branch. The metric--Maxwell entries contain at most one derivative of $a_b$, and the algebraic $\chi$ perturbation enters the metric and $W$ equations without derivatives. For an auxiliary amplitude $z=\delta\chi=O(1)$, the Maxwell equation gives $a_b=O(\lambda^{-1})$. The scalar wave block then gives $\delta W=O(\lambda^{-2})$, while the harmonic-gauge Einstein equation is sourced at order $O(1)$ by the first derivatives of $a_b$ and gives $\delta\widetilde g_{ab}=O(\lambda^{-2})$:
\begin{equation}
 a_b=O(\lambda^{-1}),\qquad
 \delta W=O(\lambda^{-2}),\qquad
 \delta\widetilde g_{ab}=O(\lambda^{-2}).
 \label{eq:app-amplitude-orders}
\end{equation}
Substituting these amplitudes into the retained $\chi$ constraint changes it only at subleading order; no inverse power of $T$ is involved. The Maxwell field strength remains of order one because $\delta F_{ab}=i(q_aa_b-q_ba_a)=O(1)$. Background gradients of $W$ introduce first-derivative metric--scalar mixing and leave the displayed orders unchanged.

To compare the two frames, let $q^2=g^{ab}q_aq_b$ and $j^b=q_aJ^{ab}$ use the Jordan metric, as in Sect.~\ref{Sec:Hyperbolicity}. Their Einstein-frame counterparts obey
\begin{equation}
 \widetilde q^2=\frac{q^2}{W},\qquad
 \widetilde J^{ab}=\frac{J^{ab}}{W^2},\qquad
 \widetilde j^2=\frac{j^2}{W^3},\qquad
 \widetilde T=\frac{T}{W^2}.
 \label{eq:app-conformal-factors}
\end{equation}
Hence $h\widetilde T\widetilde q^2+4\widetilde j^2=W^{-3}(hTq^2+4j^2)$. We express the reduced equations below in Jordan-frame contractions, absorbing nonzero conformal factors into the amplitudes. Write $a^b=\lambda^{-1}\bar a^b$ and $\widehat j^b=\widehat q_aJ^{ab}$ with $J$ defined in \eqref{eq:covariant-J}. The leading Maxwell equations, including their longitudinal component, and the algebraic $\chi$ equation become
\begin{equation}
 h\widehat q^2\bar a^b-i\widehat j^bz=0,
 \qquad Tz-4i\widehat j^b\bar a_b=0.
 \label{eq:app-reduced-system}
\end{equation}
Since $\widehat q_b\widehat j^b=0$, the first equation automatically satisfies the Lorenz condition and Gauss's law. Eliminating only the non-null Maxwell wave block gives
\begin{equation}
 \bar a^b=\frac{i\widehat j^b}{h\widehat q^2}z,
 \qquad hT\widehat q^2+4\widehat j^2=0.
 \label{eq:app-Schur}
\end{equation}
This is the covariant polynomial \eqref{eq:covariant-root}; no inverse of $T$ has appeared. Metric and scalar feedback enter the reduced equation at relative order $\lambda^{-1}$ or smaller, so its simple roots are unchanged at leading order. The same calculation applies for $T=0$.

The same reduction determines the rank and polarization of the characteristic root. On the four Maxwell components and $z$, the leading block is
\begin{equation}
 \mathsf M(\widehat q)=
 \begin{pmatrix}
 h\widehat q^2\mathbf1_4&-i\widehat j^b\\
 -4i\widehat j_b&T
 \end{pmatrix}.
 \label{eq:app-constraint-matrix}
\end{equation}
For $h>0$ and $\widehat q^2\ne0$, the Maxwell block is invertible. Its Schur complement vanishes exactly on \eqref{eq:app-Schur}, where $\mathsf M$ consequently has a one-dimensional kernel with $z\ne0$ and $\bar a^b=i\widehat j^b z/(h\widehat q^2)$. The contraction $\widehat q\cdot\bar a=0$ satisfies the longitudinal Maxwell equation as well as the transverse equations. Since $\widehat j\ne0$ is orthogonal to the non-null $\widehat q$, the field strength $\widehat q\wedge\bar a$ is nonzero. The kernel is thus a physical electromagnetic perturbation coupled to the curvature auxiliary field. At $T=0$ the Schur condition reduces to $\widehat j^2=0$, and the same one-dimensional kernel remains.

It remains to check the subsidiary constraints of the Einstein and Maxwell gauge choices. Let $E_g^{ab},E_W,E_\chi,E_A^a$ denote the Euler--Lagrange expressions of the Einstein-frame auxiliary action, normalized as variational derivatives with respect to $\widetilde g_{ab},W,\chi,A_a$. Gauge invariance gives the off-shell identities
\begin{equation}
 \widetilde\nabla_a E_A^a\equiv0,\qquad
 2\widetilde\nabla_aE_g^{ab}-E_W\widetilde\nabla^bW-E_\chi\widetilde\nabla^b\chi-E_A^a\widetilde F^b{}_a\equiv0.
 \label{eq:app-noether-identities}
\end{equation}
The second identity follows by writing an infinitesimal diffeomorphism of the potential as $\delta_\xi A_a=\xi^bF_{ba}+\widetilde\nabla_a(\xi^bA_b)$; the total derivative term drops by the first identity. Linearize the identities on a background solution, add Lorenz and harmonic gauge-fixing terms to the respective equations, and denote the gauge violations by $\mathfrak L=\widetilde\nabla_a a^a$ and $\mathfrak H_b=\widetilde\nabla^a\overline{\delta\widetilde g}_{ab}$, up to the standard lower-order background connection terms. When the auxiliary scalar and gauge-fixed field equations hold, their subsidiary principal parts are
\begin{equation}
 \sigma_{\widehat q}(E_{\rm sub}^{A})\mathrel{\propto}h\widetilde q^2\mathfrak L,
 \qquad \sigma_{\widehat q}(E_{\rm sub}^{g})_b\mathrel{\propto}\widetilde q^2\mathfrak H_b.
 \label{eq:app-subsidiary}
\end{equation}
For $h,W>0$, the proportionality factors are nonzero. Because $\widetilde q^2\ne0$, the subsidiary equations set the leading Lorenz and harmonic gauge violations to zero. The kernel constructed above therefore satisfies the physical Maxwell and Einstein constraints. The auxiliary equations $R=\chi$ and $W=f_\chi$ complete the equivalence with the original theory, including at $T=0$.

\begin{proposition}[Persistence of a physical non-null root]\label{prop:full-root}
At a smooth point with $h,W>0$, suppose $\widehat q^2\ne0$, $\widehat j\ne0$ and $\mathcal P(\widehat q)=0$ is simple as a root in the frequency variable. For all sufficiently large tangential wave numbers $k$, the complete frozen gauge-fixed linearized equations have a frequency $\omega(k)=k\widehat\omega+O(1)$ and a polarization with nonzero electromagnetic field strength. This holds whether or not $T$ vanishes. The corresponding non-null solution is not a gauge or subsidiary-constraint mode.
\end{proposition}
\begin{proof}
Fix the spatial direction and set $\lambda=|k|$. On $\widehat q^2\ne0$, the harmonic metric and scalar blocks are invertible. So is the Maxwell block on the Lorenz subspace. Their Schur complement is analytic in $\lambda^{-1}$ and has, by \eqref{eq:app-Schur}, the leading scalar factor $\mathcal P(\widehat q)/(h\widehat q^2)$. The implicit-function theorem at a simple root supplies $\widehat\omega(\lambda)=\widehat\omega+O(\lambda^{-1})$. The corresponding Maxwell amplitude is proportional to $\widehat j^b$, which is nonzero and orthogonal to the non-null covector. Hence $\widehat q\wedge\bar a\ne0$ and the field strength is nonzero.

The rank-one kernel and the Noether identities \eqref{eq:app-constraint-matrix}--\eqref{eq:app-subsidiary} show explicitly that neither Maxwell nor gravitational subsidiary modes contribute a non-null root here. The auxiliary equations hold by construction. The root therefore belongs to the original constrained system, including at $T=0$.
\end{proof}

For angular propagation $\widehat q_s=0$ and $c_\perp^2=M/L<0$, the two leading roots are
\begin{equation}
 \widehat\omega_\pm=\pm i\sqrt{-c_\perp^2},\qquad
 \widehat q^2=1-c_\perp^2>0.
\end{equation}
They are simple when $L\ne0$, so Proposition~\ref{prop:full-root} proves Eq.~\eqref{eq:frozen-growth}. The negative-cone witnesses in Theorem~\ref{thm:chargedangular} have precisely these properties. The ordinary photon and gravitational factors remain necessary physical factors; their hyperbolicity directions are incompatible with this extraordinary factor by Theorem~\ref{thm:no-common-normal}.

The exact $T=0$ lift in \eqref{eq:Tzeroresponses}--\eqref{eq:Tzerogeometry} gives an especially direct check. With $x=R_0-R$ and $\ell_0>0$, the background quantities satisfy
\begin{equation}
 T=0,\quad W=2,\quad
 \alpha=\frac{2\beta b\ell_0x^2}{x^2+\ell_0^2},\quad
 \beta_1=-\frac{\beta b x(x^2+3\ell_0^2)}{x^2+\ell_0^2}.
 \label{eq:app-Tzero-check}
\end{equation}
Both $\alpha$ and $\beta_1$ are nonzero at every positive radius of this branch. The auxiliary scalar mixing $m=(W-2)/W^3$ vanishes, and the non-null part of \eqref{eq:auxmatrix} reduces directly to $\alpha^2\omega^2+\beta_1^2k^2=0$. Its field-strength polarization is given by \eqref{eq:physicalnonnullvector}, with squared speed \eqref{eq:Tzeroexactinstability}. Thus the rank-changing curvature Hessian does not remove the characteristic in this exact solution.

The reduction identifies physical non-null roots of the frozen linearized equations. A global mode spectrum on a variable-coefficient black-hole background is a separate problem; the local common-hyperbolicity obstruction is already fixed by the characteristic factors.

\section{An analytic \texorpdfstring{$\Phi(R)$}{Phi(R)} counterexample}\label{app:phi-example}
The construction summarized in Sect.~\ref{Sec:Validity} is given here with its field-equation and principal-sign checks. The following formulae use a fixed action and are exact on the stated positive-radius interval.

Fix $P=1$, choose length units, and set $u=r^2$. Take
\begin{equation}
 \begin{split}
 W&=1+u/2,\qquad C=(1+u)^{1/4},\\
 B&=1-u+\frac{13}{20}u^2-\frac{17}{32}u^3,\qquad
 A=C^2B,
 \end{split}
 \label{eq:Phi-example-metric}
\end{equation}
with $F=\sin\vartheta\,\dd\vartheta\wedge\dd\varphi$. Define the action parametrically by
\begin{align}
 R(u)&=\frac{11305u^4+13280u^3+448u^2+2880u+2880}
 {320(1+u)^2},\label{eq:Phi-example-R}\\
 h(R(u))&=\frac{u^4(80-15892u-22520u^2-8075u^3)}
 {2560(1+u)^2},\label{eq:Phi-example-h}\\
 \Phi(R(u))&=\frac{1615u^5+26380u^4+25364u^3-928u^2
 +11520u+11520}{1280(1+u)^2}.
 \label{eq:Phi-example-Phi}
\end{align}
Since $R(0)=9$ and $R_u(0)=-9$, the analytic inverse-function theorem defines single-valued analytic $h(R)$ and $\Phi(R)$ near $R=9$. The coefficients specify one fixed action. Both metric functions are analytic in $r^2$, with $A(0)=B(0)=1$, and $(B^{-1}-1)/r^2$ is analytic; the metric therefore extends analytically in Cartesian coordinates. The magnetic two-form retains its nonzero conserved flux and is singular on the filled ball.

For completeness, the inverse construction verifies all equations. The functions in \eqref{eq:Phi-example-metric} obey
\begin{equation}
 W''-\frac{C'}C W'-\frac{2C'}{rC}W=0.
\end{equation}
The other two independent metric combinations give
\begin{equation}
 \begin{split}
 \frac{h}{r^4}&=\frac12\left[
 W(G^\vartheta{}_\vartheta-G^t{}_t)
 +B\left(\frac{A'}{2A}-\frac1r\right)W'\right],\\
 \Phi&=\frac12(WR+3\Box W).
 \end{split}
 \label{eq:Phi-inverse-construction}
\end{equation}
Substitution yields exactly \eqref{eq:Phi-example-R}-- \eqref{eq:Phi-example-Phi}. The difference equation, angular-minus- temporal equation and trace imply all three independent metric equations. The Bianchi identity then requires
\begin{equation}
 \frac{\dd\Phi}{\dd r}-\frac2{r^4}\frac{\dd h}{\dd r}
 =W\frac{\dd R}{\dd r}.
\end{equation}
Direct differentiation of the rational functions verifies this identity and hence $\Phi_R-2h_R/r^4=W$. Finally, $\dd F=0$ and $\dd(h{}^\star F)=0$ hold identically: the latter two-form is a radial function multiplying $\dd t\wedge\dd r$. Thus the reconstructed action satisfies both the gravitational and electromagnetic equations.

Introduce
\begin{align}
 p(u)&=80-15892u-22520u^2-8075u^3,\notag\\
 d(u)&=11305u^4+29250u^3+19920u^2-992u-1440,\notag\\
 t(u)&=62985u^3+142120u^2+80516u-3200.
\end{align}
Then
\begin{equation}
 R_u=\frac{d(u)}{160(1+u)^3},\qquad
 T=\frac{40(1+u)^4t(u)}{d(u)^2}.
\end{equation}
For $0\le u\le10^{-3}$, bounding each positive monomial by its endpoint value gives $B>0$, $p>64$, $d<-1439$ and $t<-3119$. Consequently $h>0$, $W>0$, $R_u\ne0$ and $T<0$ at every positive point of this interval.

Since $e=0$, the photon kinetic coefficient is $h$ and
\begin{equation}
 c_\perp^2=1+\frac{4h_R^2}{u^2hT}.
 \label{eq:Phi-example-speed}
\end{equation}
To verify its positive sign explicitly, write
\begin{align}
 s(u)={}&560768375u^8+4127552400u^7+12880725015u^6
 \notag\\
 &+21845401480u^5+21255236052u^4+11260874088u^3
 \notag\\
 &+2544854624u^2+3015840u-76800.
\end{align}
Exact simplification gives
\begin{equation}
 c_\perp^2=\frac{2s(u)}{(1+u)^2p(u)t(u)}.
\end{equation}
The sum of the positive monomials of $s$ is less than $6000$ on $[0,10^{-3}]$, so $s<0$ and $c_\perp^2>0$. The derivative $h_R$ is nonzero for $0<u\le10^{-3}$: its nontrivial numerator is $40375u^4+146605u^3+182796u^2+79300u-320<0$. Equation~\eqref{eq:Phi-example-speed} therefore also gives $c_\perp^2<1$.

The limiting coefficients are
\begin{equation}
 \begin{split}
 h&=u^4/32+O(u^5),\qquad \Phi_R=1+17u/36+O(u^2),\\
 \Phi_{RR}&\longrightarrow-17/324,\qquad
 T\longrightarrow-5/81,\qquad c_\perp^2\longrightarrow3/5.
 \end{split}
\end{equation}
The extraordinary covector form is $-\omega^2+k_r^2+c_\perp^2k_\perp^2$, and the other photon, scalar and tensor cones are luminal. They share the static time as a hyperbolic normal. The scalar kinetic term is $-3(\widetilde\nabla\log W)^2/2$ and the photon kinetic coefficient is $h>0$; the negative auxiliary Hessian $T$ therefore corresponds to no kinetic ghost in this local principal system. Lower-order mass terms can be studied independently.

The positivity holds at each positive radius in the stated interval, with $h\to0$ at its endpoint. Indeed, $W-\Phi_R=O(r^2)$ while $\Phi_{RR}$ and $2h_{RR}/r^4$ both approach finite values. Their competition in \eqref{eq:Phi-principal-shift} is what permits the subluminal extraordinary cone.

\acknowledgments

\paragraph{Use of generative AI.}
OpenAI ChatGPT was used to assist with literature organization, manuscript preparation, and the exploration and checking of mathematical arguments. The author reviewed the AI-assisted material and takes full responsibility for the manuscript.

\paragraph{Funding.}
No funds, grants, or other support were received.

\paragraph{Competing interests.}
The author has no relevant financial or non-financial interests to disclose.

\paragraph{Data availability.}
No datasets were generated or analyzed during this study.

\bibliographystyle{JHEP}
\bibliography{biblio}

\end{document}